\documentclass[aps,onecolumn,showpacs]{revtex4-1}
\usepackage{graphicx}
\usepackage{amsmath,amsfonts}
\usepackage{bm}
\usepackage[T1]{fontenc}
\usepackage{xcolor}
\def\fpint{\rlap{$\,\hspace{0.3785pt}\times$}\int} % cross integral
\def\cpvint{\rlap{$\,\hspace{0.3785pt}-$}\int} % CPV integral
\def\rd{{\rm d}}
\def\ci{{\rm i}}
\def\ce{{\rm e}}
\def\be{\begin{equation}}
\def\ee{\end{equation}}
\def\LLE#1#2{(${\rm L}^{#1}_{#2}$)}
\def\pdv#1#2{\frac{\partial #1}{\partial #2}}
\def\pdvt#1#2{\partial #1/\partial #2}
\def\Ac{{\mathcal A}}
\def\Bc{{\mathcal B}}
\def\Kc{{\mathcal K}}
\def\fpint{\rlap{$\,\hspace{1.0pt}\times$}\int} % cross integral
\def\cpvint{\rlap{$\,\hspace{1.0pt}-$}\int} % CPV integral
\def\Kt{\widetilde{K}}
\def\Mt{\widetilde{M}}
\def\ut{\tilde{u}}
\def\gt{\tilde{g}}
\def\gte{\gt_{{\rm e}}}
\def\gto{\gt_{{\rm o}}}
\def\Real{{\rm Re}}
\def\Imag{{\rm Im}}
\def\ve{\varepsilon}
\def\newint{\rlap{\ $\to$}\int}
\def\eqn{Eq.}
\def\eqns{Eqs.}

\begin{document}

% Add a serial/Oxford comma by default.
%\newcommand{\reflastconjunction}{, and~}

% Used for creating new theorem and remark environments

% Sets running headers as well as PDF title and authors
%\headers{Love--Lieb integral equations}{L. Farina, G. Lang and P. A. Martin}

\title{Love--Lieb integral equations: applications, theory, approximations, and computations}
%\funding{This work was funded by the Fog Research Institute under contract no.~FRI-454.}}}

% Authors: full names plus addresses.
%\author{Dianne Doe\thanks{Imagination Corp., Chicago, IL 
%  (\email{ddoe@imag.com}, \url{http://www.imag.com/\string~ddoe/}).}
%\and 
%Paul T. Frank\thanks{Department of Applied Mathematics, Fictional University, Boise, ID 
%  (\email{ptfrank@fictional.edu}, \email{jesmith@fictional.edu}).}
%\and 
%Jane E. Smith\footnotemark[3]}

\author{Leandro Farina}
\affiliation{Institute of Mathematics and Statistics, Federal University of Rio Grande do Sul, Porto Alegre, Brazil.}

\author{Guillaume Lang}
\affiliation{CNRS, LPMMC, F-38000 Grenoble, France.}

\author{P.A. Martin}
\affiliation{Department of Applied Mathematics and Statistics, Colorado School of Mines, Golden, CO 80401, USA.}

%\usepackage{amsopn}
%\DeclareMathOperator{\diag}{diag}

% Optional PDF information
%\ifpdf
%\hypersetup{
%  pdftitle={Love--Lieb integral equations},
%  pdfauthor={L. Farina, G. Lang and P. A. Martin}
%}
%\fi

% The next statement enables references to information in the
% supplement. See the xr-hyperref package for details.

% \externaldocument{ex_supplement}

% FundRef data to be entered by SIAM
%<funding-group specific-use="FundRef">
%<award-group>
%<funding-source>
%<named-content content-type="funder-name"> 
%</named-content> 
%<named-content content-type="funder-identifier"> 
%</named-content>
%</funding-source>
%<award-id> </award-id>
%</award-group>
%</funding-group>

% REQUIRED
\begin{abstract}
This paper is concerned mainly with the deceptively simple integral equation
\[
   u(x) - \frac{1}{\pi}\int_{-1}^{1} \frac{\alpha\, u(y)}{\alpha^2+(x-y)^2} \, \rd y = 1, \quad -1 \leq x \leq 1,
\]
where $\alpha$ is a real non-zero parameter and $u$ is the unknown function. 
This equation is classified as a Fredholm integral equation of the second kind with a continuous kernel. 
As such, it falls into a class of equations for which there is a well developed theory. 
The theory shows that there is exactly one continuous real solution $u$. 
Although this solution is not known in closed form, it can be computed numerically, using a variety of methods.
All this would be a curiosity were it not for the fact that the integral equation arises in several contexts in classical and quantum physics.
We review the literature on these applications, survey the main analytical and numerical tools available, 
and investigate methods for constructing approximate solutions. 
We also consider the same integral equation when the constant on the right-hand side is replaced by a given function.
\end{abstract}

% REQUIRED
%\begin{keywords}
%Love's integral equation, Lieb's integral equation,
%Gaudin's integral equation
%\end{keywords}

\maketitle

% \tableofcontents

\section{Introduction}

It is well known that one partial differential equation can appear in several models of disparate physical phenomena:
one thinks immediately of the three classic examples, Laplace's equation, the wave equation and the diffusion equation.
Actually, it is the case for certain integral equations as well. 
In this review paper, we consider one such family of integral equations, usually associated with the names of E.~R.~Love and E.~H.~Lieb (although other names could stake a claim, as we shall see).
The simplest {\it Love--Lieb equation\/} reads
\be\label{LLeqn}
u(x)\pm \frac{1}{\pi} \int_{-1}^{1} \frac{\alpha\, u(y)}{\alpha^2+(x-y)^2} \, \rd y = 1, \quad -1 \leq x \leq 1,
  \tag{\mbox{${\rm L}^\pm_1$}}
\ee
where $\alpha$ is a positive real parameter and $u$ is the unknown function. 
Let us clarify our notation. 
The superscript $\pm$ in the label (\ref{LLeqn}) refers to the sign before the integral, and the subscript 1 refers to the function on the right-hand side. 
Later, we shall encounter \LLE{\pm}{g} when the right-hand side is replaced by $g(x)$ and, in particular, \LLE{\pm}{x} when $g(x)=x$. 
As the solution $u$ depends on the value of $\alpha$, we shall write $u(x;\alpha)$ when that dependence matters.

We could have written (\ref{LLeqn}) as a single equation just by allowing $\alpha$ to be negative as well as positive. 
However, it turns out that the solution does not behave continuously as $\alpha$ passes through zero, and so for some purposes 
it is more convenient to be able to identify two distinct equations, \LLE{+}{1} and \LLE{-}{1}.
These two integral equations also have distinct applications (see \ref{contextsec}).
On the other hand, the distinction between \LLE{+}{1} and \LLE{-}{1} is largely irrelevant when it comes to solvability (\ref{techniques}) or choice of numerical method (\ref{numerics}).

Let us outline two physical problems leading to (\ref{LLeqn}).
In his 1949 paper \cite{Love1949}, Love derived (\ref{LLeqn}) in the context of an electrostatic problem: 
determine the potential field about two identical charged coaxial circular discs. 
This structure is called a {\it circular plate capacitor}.
The parameter $\alpha=d/R$, where the discs have radius $R$ and $d$ is the distance between them.
Equation \LLE{+}{1} is appropriate when the discs are equally charged, whereas \LLE{-}{1} holds when the discs are oppositely charged. 
The solution $u$ of \ref{LLeqn} is an auxiliary function: the exact electrostatic potential field $\phi$ is given as a certain integral of $u$.
In particular, the capacitance is proportional to $C=\int_{-1}^1u(x)\,\rd x$.
The problem of determining $\phi$ and $C$ has a long history, stretching back to the middle of the 19th century;
for historical remarks and references, see \cite[section~8.1]{Sneddon66} and \cite{Kuester1987}.
Love proved that each of (\ref{LLeqn}) has exactly one solution $u$, which is an even, real-valued, continuous function on the interval $[-1,1]$. 
However, the exact solution is not known in closed form.
For more details, see Sec.~\ref{CPCsec}. 
Actually, in accord with Stigler's Law of Eponymy, ``Love's equation'' had already been derived in 1910 by Hafen \cite[section~3, p.~529, \eqn~(10)]{Hafen10}
for the same capacitor problem.

The integral equations \ref{LLeqn} also appear in condensed matter physics, more specifically, in the context of certain quantum integrable models.
These models describe a one-dimensional gas of identical particles.
The Lieb--Liniger model leads to the {\it Lieb equation \/} \LLE{-}{1}; see \cite[\eqn~(3.18)]{LiebLiniger1963}. 
In this model, the particles are spinless bosons and their interaction is repulsive.
In the Yang--Gaudin model \cite{Yang1967, Gaudin1968}, the particles are spin-$\frac{1}{2}$ fermions;
their interaction can be repulsive \cite{Yang1967} or attractive \cite{Gaudin1968}.
Here, both cases lead to coupled integral equations 
but, when the total spin is zero and the interaction is attractive, a single integral equation is obtained, namely \LLE{+}{1}; we call this the {\it Gaudin equation\/} \cite[\eqn~(7)]{Gaudin1967}.
In both models, the parameter $\alpha$ is related to the strength of the two-body interaction, and the ground-state energy density can be calculated using certain integrals of $u$ in the thermodynamic limit.
For more details, see \ref{QIMsec}. 
The observation that the Lieb and Gaudin integral equations are the same as those studied by Love \cite{Love1949}
and Sneddon \cite{Sneddon66} was made by Gaudin in his 1968 thesis \cite{Gaudin1968}; see also \cite{Gaudin1971}.

In what follows, we shall also be concerned with a generalization of (\ref{LLeqn}),
\be \label{GLLu}
   u(x) \pm \frac{1}{\pi}   \int_{-1}^{1} \frac{\alpha\, u(y)}{\alpha^2+(x-y)^2} \, \rd y = g(x), \quad -1 \leq x \leq 1,
     \tag{\mbox{${\rm L}^\pm_g$}}
\ee
where $g(x)$ is a given function and, as before, $\alpha$ is a positive real constant.
This is a Fredholm integral equation of the second kind with a continuous symmetric difference kernel, $K(x-y)$, where $K(x)=(\alpha/\pi)(\alpha^2+x^2)^{-1}$
is known as the {\it Cauchy distribution\/} or the {\it Lorentzian function}.
When $g(x)=1$ in (\ref{GLLu}), we recover (\ref{LLeqn}). 
Henceforth, we shall refer to (\ref{GLLu}) as the {\it generalized Love--Lieb\/} equation. 
This formulation, with an arbitrary right-hand side function, allows us to account for variants of the Love--Lieb equation that emerge in a wide range of seemingly unrelated fields of physics and mathematics.

Recalling Stigler's Law again, we note that an early derivation of \LLE{+}{g} was already given by Hulth\'en in his 1938 thesis on anti\-ferro\-magnetic properties of crystals; see \cite[\eqn~(III,\,58)]{Hulthen38}. 
His derivation led to \LLE{+}{g} with $g(x)=(\alpha^2+4x^2)^{-1}$ (after some simple scaling).

Although the exact solution of the Love--Lieb equation (\ref{LLeqn}) is not known in closed form, efforts to solve it have stimulated the development of many mathematical and numerical methods.
Some of these will be discussed below. 
On the other hand, by inserting specific functions $u$ (such as monomials or orthogonal polynomials) into the left-hand side of (\ref{GLLu}), one can compute $g$; this trivial observation is useful when the aim is to test numerical methods; see Sec.~\ref{ExpMethSec} for details.

The structure of this paper is as follows.
In Sec.~\ref{contextsec}, we survey physics problems involving the Love--Lieb equation or generalizations thereof. 
Two main contexts are outlined: classical physics problems involving coaxial circular discs, and quantum physics problems involving one-dimensional models. 
Other types of applications are briefly discussed too. 
Section \ref{contextsec} may be omitted by readers interested solely in mathematical aspects; these are the focus of the remainder of the paper.
Section \ref{techniques} contains a summary of theoretical results for the Love--Lieb equation, 
Sec.~\ref{numerics} summarizes the main numerical methods that can be used to solve it, 
and Sec.~\ref{LLapproxSec} is devoted to analytical approximations.
Appendix \ref{loveinftySec} discusses (\ref{GLLu}) over an infinite range,
\[
   u(x) \pm \frac{1}{\pi}   \int_{-\infty}^{\infty} \frac{\alpha\, u(y)}{\alpha^2+(x-y)^2} \, \rd y = g(x), \quad -\infty \leq x \leq \infty,
\]
an equation that can be treated (formally, at least) using Fourier transforms.
We make some concluding remarks in Sec.~\ref{cclSec}.

%%%%%%%%%%%%%%%%%%%%%%%%%%%%%%%%%%%%%%%%%
\section{Applications of Love--Lieb integral equations}\label{contextsec}
%%%%%%%%%%%%%%%%%%%%%%%%%%%%%%%%%%%%%%%%%

\subsection{Potential problems involving two coaxial discs}

We consider axisymmetric boundary-value problems for a potential function $\phi(r,z)$, where $r$ and $z$ are cylindrical polar coordinates.
There are two coaxial discs of radius 1, one in the plane $z=0$ and one in the plane $z=\alpha$.  
We solve Laplace's equation in three dimensions, $\nabla^2\phi=0$, outside the discs together with a far-field condition, $\phi=O(R^{-1})$ as $R=\sqrt{r^2+z^2}\to\infty$, and boundary conditions on both discs.

\subsubsection{Circular plate capacitor}\label{CPCsec}

For electrostatic (capacitor) problems, $\phi$ is prescribed on each disc.
We take $\phi=1$ on the lower disc (at $z=0$) and $\phi=\pm 1$
on the upper disc; the solution with $\phi=+1$ ($\phi=-1$) 
on the upper disc corresponds to the case of ``equally charged discs''  (``oppositely charged discs'' respectively). 
The solutions for these two problems are given in Sneddon's book \cite{Sneddon66}. 
The basic physical quantity to be computed is the charge density $\sigma$. On the lower disc, we have
\begin{align}
    \sigma(r)&=-\frac{1}{4\pi} \left( 
    \left.\pdv{\phi}{z}\right|_{z=0^+}
    -\left.\pdv{\phi}{z}\right|_{z=0^-}\right) \nonumber\\
    &=\frac{1}{\pi^2}\int_0^\infty J_0(kr)\int_0^1 u(t)\, k \cos(kt) \,\rd t\,\rd k, \label{sigintA}
\end{align}
using \cite[\eqns\ (1.1.4), (8.1.12) and (8.1.14)]{Sneddon66}, where $J_0$ is a Bessel function and $u$ solves (\ref{LLeqn}). The inner integral is
\[
  \int_0^1 u(t)\,\frac{\rd}{\rd t}\left(\sin(kt)\right)\,\rd t
  =u(1)\sin(k)-\int_0^1u'(t)\sin(kt)\,\rd t.
\]  
Substituting in (\ref{sigintA}), we can change the order of integration followed by use of \cite[\eqn~(2.1.14)]{Sneddon66}, giving $\sigma(r)=0$ for $r>1$ (as expected) and
\begin{align}
  \pi^2\sigma(r)&=\frac{u(1)}{\sqrt{1-r^2}} 
  -\int_r^1 \frac{u'(t)\,\rd t}{\sqrt{t^2-r^2}}\label{sigfone}\\
  &=-\frac{1}{r}\,\frac{\rd}{\rd r}\int_r^1 
  \frac{t u(t)\,\rd t}{\sqrt{t^2-r^2}}
\end{align}
for $0\leq r<1$. Inverting, using \cite[\eqn~(2.3.8)]{Sneddon66}, yields
\be
  u(x)=2\pi\int_x^1 \frac{r\sigma(r)\,\rd r}{\sqrt{r^2-x^2}}.
  \label{usigint}
\ee
This known formula \cite[\eqn~(1.9)]{LeppingtonL70},
\cite[\eqn~(7)]{Tracy2016} is useful because it relates $u$ to a physical quantity, $\sigma$, which has known properties. 
For example, $\sigma(r)$ behaves as an inverse square-root as $r\to 1$; see the first term on the right-hand side of (\ref{sigfone}). 
Also, as $u(t)$ is even, (\ref{sigfone}) implies that $\sigma$ is even too.
These facts suggest expanding $\sigma(r)$ using functions of the form $(1-r^2)^{-1/2}\psi_n(r)$, where $\psi_n(r)=\psi_n(-r)$ is a polynomial.
If we try $\psi_n(r)=r^{2n}$, $n=0,1,2,\ldots$ and substitute for $\sigma$ in (\ref{usigint}), some calculation shows that $u(x)$ is a polynomial in $x^2$ of degree $n$.

A less obvious choice is  $\psi_n(r)=P_{2n}(\sqrt{1-r^2}\,)$, where $P_m$ is a 
Legendre polynomial.  These functions
(which evaluate to polynomials in $r^2$ of degree $n$)
 are useful for single-disc problems and they permit an explicit calculation of the corresponding
function $u(x)$ because of the formula \cite[p.~357]{England81}
\[
   2\pi\int_x^1 \frac{r\,P_{2n}(\sqrt{1-r^2}\,)}{\sqrt{r^2-x^2}\sqrt{1-r^2}}
   \,\rd r=\pi^2 P_{2n}(0)\,P_{2n}(x),\quad 0\leq x\leq 1.
\]
This motivates the use of Legendre polynomials to approximate $u(x)$.
We shall return to this topic in Sec.~\ref{ExpMethSec}.

The total charge on the lower disc is
\be
  \int_0^1\int_{-\pi}^\pi \sigma(r)\,r\,\rd\theta\,\rd r
  =\frac{1}{\pi}C(\alpha)
  \quad\mbox{with}\quad
  C(\alpha)=\int_{-1}^1u(x;\alpha)\,\rd x.\label{CCap}
\ee
Asymptotic approximations of $C(\alpha)$ for small gaps ($\alpha\ll 1$) and for large gaps ($\alpha\gg 1$) have been found using various methods.
For surveys, see, for example, \cite[section~8.1]{Sneddon66} and  \cite{Kuester1987, Soibelman1996, Reichert2020}.

The capacitor problem with discs of unequal radii leads to a pair of coupled integral equations of Love type \cite{Collins61, Paffuti16}.
For several coaxial discs, see~\cite{Fabrikant87}.

\subsubsection{Potential flow past rigid discs}

For potential flow past rigid discs, $\pdvt{\phi}{z}$ is
prescribed on each disc. There are two basic problems, both with $\pdvt{\phi}{z}=1$ on the lower disc.
One problem has $\pdvt{\phi}{z}= 1$ on the upper disc, the other has $\pdvt{\phi}{z}= -1$ there.
The basic unknown is $[\phi](r)$, the jump in $\phi(r,z)$ across the lower disc, defined by
\[
  [\phi](r)=\phi(r,0^+)-\phi(r,0^-).
\]
It is shown in \cite{martin1997} that if we write
\begin{equation}
   [\phi](r)= -\frac{4}{\pi} \int_{r}^{1} \frac{u(t)\,\rd t}{\sqrt{t^2 - r^2}}, \label{phijump}
\end{equation}
then $u$ solves a special case of (\ref{GLLu}), namely,
\be \label{LiebTwo}
   u(x) \pm \frac{1}{\pi}   \int_{-1}^{1} \frac{\alpha\, u(y)}{\alpha^2+(x-y)^2} \, \rd y = x, \quad -1 \leq x \leq 1.
     \tag{\mbox{${\rm L}^\pm_x$}}
\ee
The solution $u$ of (\ref{LiebTwo}) is odd, real, and continuous on the interval $[-1,1]$. 
The integral equations (\ref{LiebTwo}) can also be extracted from a paper by Collins \cite{Collins61}; 
this paper also derives coupled integral equations of Love--Lieb type for discs of unequal radii.

Equation (\ref{phijump}) can be inverted, using \cite[\eqn~(2.3.8)]{Sneddon66},
\[
  u(x)=\frac{1}{2}\,\frac{\rd}{\rd x}\int_x^1 \frac{r [\phi](r)}{\sqrt{r^2-x^2}}\,\rd r, 
\]
and this could be used to generate expansions for $u$ using known properties of $[\phi]$; for example, $[\phi](r)$ has a square-root zero as $r\to 1$.

In the context of irrotational flow of an inviscid incompressible fluid, the force on the lower disc can be expressed in terms of the added mass
\cite[\eqn~(6.2{\it a\/})]{atkinson2017},
\[
\Ac(\alpha) = -2\pi \int_0^1 [\phi](r) \, r\, \rd r = 8\int_0^1 u(x;\alpha)\,x\,\rd x,
\]
where $u$ is once again the solution of (\ref{LiebTwo}).
Analytical approximations of  $\Ac(\alpha)$ for $\alpha\ll 1$ are obtained in~\cite{atkinson2017}.

Cooke \cite[p.~108]{Cooke1956} first derived (\ref{LiebTwo}) in 1956 for the problem of two discs rotating slowly in a viscous fluid (Stokes flow), with equal, or equal and opposite, angular velocities; see also \cite{Cooke1958}. For an approximation to the torque on each disc
when $\alpha\ll 1$, see~\cite{Hutson64}.

Suppose next that $z=0$ is the mean free surface of deep water. 
Small-amplitude water waves are generated by the vertical oscillations of a rigid disc submerged at a depth of $\alpha/2$.  
The motion can be calculated by solving a generalization of (\ref{LiebTwo}) \cite{martin1997},
\begin{equation}
  u(x)
  -\frac{1}{\pi}\int_{-1}^1 \frac{\alpha\,u(y) }{\alpha^2 + (x-y)^2}\,\rd y
  -\frac{2K}{\pi}\int_{-1}^1 u(y) \,\Phi(x-y,\alpha)\, \rd y
  = x,   \quad -1\leq x\leq 1,             \label{eq:lovegen}
\end{equation}
where $K=\omega^2/g$ is the wave number, $\omega$ is the frequency, $g$ is the acceleration due to gravity, $\Phi$ is a two-dimensional wave-source potential given by
\[
  \Phi(X,Y)=\int_0^\infty \ce^{-kY}\cos(kX)\frac{\rd k}{k-K},                         
\]
and the integration path is indented below the pole of the integrand at $k = K$. 
As before, the discontinuity in $\phi$ across the disc, $[\phi]$, is given by (\ref{phijump}).  
Approximations for $\alpha\ll 1$ (meaning that the disc is very close to the free surface) are developed in~\cite{Farina2010}.

\subsection{Quantum integrable models}\label{QIMsec}

Quantum integrable models are a class of one-dimensional models that are exactly solvable by the Bethe ansatz \cite{Bethe1931}.
In the thermodynamic limit, the coupled Bethe ansatz equations that describe them reduce, in some cases, to a single Love--Lieb equation. 
This phenomenon occurs in the continuum (with the Lieb--Liniger and Yang--Gaudin models) and on the lattice (with the Heisenberg model).

\subsubsection{The Lieb--Liniger model}

The Lieb--Liniger model \cite{LiebLiniger1963} describes a one-dimensional gas of identical spinless bosons interacting through a contact potential. 
Proposed in 1963 as a generalization of the Tonks--Girardeau gas of hard-core bosons \cite{Girardeau1960}, it is arguably the simplest (conceptually), as well as the most studied non-trivial quantum integrable model in the continuum. 
The quantity  $u(x;\alpha)/(2\pi)$ denotes the distribution of pseudo-momenta (or rapidities) at zero temperature, $x$ is the pseudo-momentum and $\alpha$ is related to the interaction strength.

The function $u(x;\alpha)$ is defined as the solution of \LLE{-}{1}, known as the {\em Lieb equation\/} in this context. It can be used to determine quantities of physical interest. 
For example, the dimensionless average ground-state energy per particle, $e(\gamma)$, is determined by eliminating $\alpha$ between
\be
\frac{2\pi\alpha}{\gamma}= \int_{-1}^1 u(x;\alpha) \, \rd x
\quad\mbox{and}\quad
e(\gamma)=\frac{\gamma^3}{2\pi\alpha^3}\int_{-1}^1 x^2 u(x;\alpha)\,\rd x,
\label{energyLL}
\ee
where $\gamma$ is the Lieb parameter,
a dimensionless coupling constant \cite{LiebLiniger1963}.
For more information, see \cite[Chapter~1]{Korepin1993}, \cite{Cazalilla2011},
\cite[Chapter~4]{Gaudin2014},  \cite{Jiang2015, Tracy2016},  \cite[Chapter~2]{Franchini2017}. 
Many ground-state observables can be computed from derivatives of $e(\gamma)$; see, for example, \cite{Lang2017} and references therein. 
Local correlation functions can be expressed as moments of $u$ in certain approaches \cite{Cheianov2006, Olshanii2017}.
In other approaches, these correlations are calculated from the solution of \LLE{-}{g} with $g(x)=x^n$, $n=1,2,\dots$ \cite{Kormos2011, Pozsgay2011}.
This happens in a certain special case in which the function $f$ defined in \cite{Kormos2011} is such that $f(p)=0$ for $|p|>B$
and $f(p)=1$ for $|p|<B$, with $B$ a constant.
% , and $f(p)$ takes a constant value for $|p|\leq B$, when correlations are evaluated at zero temperature and at equilibrium, thus yielding a generalized Love--Lieb equation.

The Love--Lieb equation  \LLE{-}{1} is also involved in the calculation of the excitation spectrum, as is \LLE{-}{x} \cite{Pustilnik2014, Ristivojevic2014}, sometimes referred to as the {\em second Lieb equation\/} in this context. 
These equations are obtained by transforming \LLE{-}{g} with more complicated right-hand side functions $g(x)$ introduced by
Lieb himself \cite{Lieb1963bis}, using a Green function (solution of \LLE{-}{g} with $g$ replaced by a Dirac delta) \cite{Reichert2019}. 
The boundary energy is another quantity of interest
 \cite{Gaudin1971, Batchelor2005bis, Reichert2019}, whose calculation also involves \LLE{-}{1} and \LLE{-}{x} \cite{Reichert2019}.

Several generalizations of the Lieb--Liniger model also involve \LLE{-}{g}.
Equation \LLE{-}{1} appears in an extension of the model to multi\-component bosons \cite{Li2003}, and a generalized Love--Lieb equation yields its excitation energy \cite{Li2003bis}. 
Equation \LLE{-}{1} also appears in an extension of the Lieb--Liniger model to anyonic statistics, but with $\alpha$ replaced by $\alpha\sec{(\kappa/2)}$, where $\kappa\in [0,4\pi]$ is an anyonic phase parameter \cite{Batchelor2006}. 
For another generalization, leading to \LLE{-}{g} with $g(x)=(1-\beta x/\alpha)^{-2}$ and a certain parameter $\beta$, see~\cite{Stouten2018}.
 
\subsubsection{The Yang--Gaudin model}

The Yang--Gaudin model is the two-compo\-nent counterpart of the Lieb--Liniger model, with bosons replaced by spin-$\frac{1}{2}$ fermions and an arbitrary total spin $S$ compatible with the individual spins \cite{Yang1967, Gaudin1967}; see \cite{Guan2013} for a review. 
It generalizes a model studied by McGuire \cite{McGuire1965, McGuire1966}, where only one spin is flipped with respect to all the others.

Interactions between these fermions can be repulsive
\cite[\eqn~(26)]{Yang1967}, \cite[\eqn~(12)]{Guan2013} or attractive \cite[\eqns~(14.16) and (14.17)]{Gaudin1968}, \cite[\eqn~(13)]{Guan2013}.
In both cases, the result is a pair of coupled integral equations of Love--Lieb type.
The attractive case reduces to a single integral equation, \LLE{+}{1}, when $S=0$ (the so-called ``balanced case''). 
To see this, start with Gaudin's coupled equations \cite{Gaudin1968}, which we write in his notation:
\begin{align}
&
1=f_1(k)+\frac{|V|}{2\pi}\int_{-q_0}^{q_0} \frac{f(q')\,\rd q'}{(k-q')^2+V^2/4}, \quad -k_1<k<k_1, \label{G68a}\\
&
1= \frac{1}{2}f(q)+\frac{|V|}{2\pi}\int_{-q_0}^{q_0} \frac{f(q')\,\rd q'}{(q-q')^2+V^2}
   +\frac{|V|}{4\pi}\int_{-k_1}^{k_1}  \frac{f_1(k)\,\rd k }{(k-q)^2+V^2/4},  \label{G68b}
\end{align}
for $-q_0<q<q_0$.
Both unknown functions, $f$ and $f_1$, are positive. Moreover, when the total spin $S=0$, $f_1$ must satisfy \cite[\eqn~(14.13)]{Gaudin1968}
$\int_{-k_1}^{k_1}f_1(k)\,\rd k=0$,
which we enforce by letting $k_1\to 0$
\cite[p.~10]{Iida2007}. In this limit, (\ref{G68a}) becomes irrelevant whereas (\ref{G68b}) reduces to \LLE{+}{1} with $\alpha=|V|/q_0$ and $u(x)=\frac{1}{2}f(q_0x)$. 
In this context, we refer to \LLE{+}{1} as {\it Gaudin's integral equation\/} \cite[\eqn~(7)]{Gaudin1967}, \cite[\eqn~(2.35{\it a\/})]{Iida2007}, \cite[\eqn~(3)]{Tracy2016bis}.
Having solved \LLE{+}{1} for $u(x;\alpha)$, the dimensionless average ground-state energy per particle, $e(\gamma)$, is determined by eliminating $\alpha$ between
\be
\frac{\pi\alpha}{2\gamma}= \int_{-1}^1 u(x;\alpha) \, \rd x
\quad\mbox{and}\quad
e(\gamma)=-\frac{\gamma^2}{4}+\frac{2\gamma^3}{\pi\alpha^3}\int_{-1}^1 x^2 u(x;\alpha)\,\rd x;
\label{energyG}
\ee
see \cite[\eqn~(4)]{Tracy2016bis},
\cite[\eqns (6) and (7)]{MarinoReis2019}.
The low-energy spin excitations can be calculated using \LLE{+}{g} and a certain $g$ \cite{Gaudin1968, Zhou2012}.

The ground state of the balanced fermionic gas is described by \LLE{+}{1}, as is the first excited state of the attractive Lieb--Liniger model \cite{Chen2010}, the so-called ``super Tonks--Girardeau gas'' \cite{Batchelor2005, Astrakharchik2005}. 
This connection was anticipated by Gaudin \cite{Gaudin1967}. 
% \fbox{-> I've replaced "mapping" by "correspondence". Is it better like this?} \fbox{How about "connection"?}
In 2004, a modified Yang--Gaudin model that bridges the Yang--Gaudin and Lieb--Liniger models was introduced \cite{Fuchs2004,Tokatly2004} as a toy model to study the crossover from a Bose--Einstein condensate to a Bardeen--Cooper--Schrieffer state (BEC--BCS crossover) in one dimension. 
Here, both signs of the Love--Lieb equation \LLE{\pm}{1} are involved \cite{Fuchs2004, Iida2007}.
 
 The generalization of the Yang--Gaudin model to fermions with arbitrary half-integer spin $s$ \cite{Sutherland1968, Takahashi1970} is sometimes called the $\kappa$-component model, where $\kappa=2s+1$ \cite{Guan2012bis}; the result is $\kappa$ coupled integral equations.
 In the infinite-spin limit ($\kappa\to\infty$), an exact mapping (infinite-spin bosonization) transforms the thermodynamics of the $\kappa$-component model with repulsive interactions into that of a single Lieb--Liniger gas \cite{Yang2011, Liu2014}. 
 As a consequence, the Lieb--Liniger model described by \LLE{-}{1} is a good approximation to multicomponent fermions with a high number of internal degrees of freedom.
 
 \subsubsection{The Heisenberg model}

 The Heisenberg model (also known as the XXX spin chain) refers to an isotropic one-dimensional chain of quantum spins with nearest-neighbor interactions \cite{Heisenberg1928}. Its solution was provided by Bethe in 1931, making use of his ansatz technique \cite{Bethe1931}.
 As already mentioned above, it was in this context that Hulth\'en first obtained a generalized Love--Lieb equation \cite[\eqn~(III,\,58)]{Hulthen38}; see also \cite{Griffiths1964} for a more comprehensive study.
 
 The analysis of the model depends on the value of the spin $s$. For $s=\frac{1}{2}$, \LLE{+}{1} and \LLE{+}{g} are involved \cite{Cabra1998, Hammar1999}. 
 For $s=-1$ (this formal case with $s<0$ can be viewed as an effective field theory of Quantum Chromodynamics), integral equations of the type \LLE{-}{1} and \LLE{-}{g} are obtained, as for arbitrary negative spin \cite{Hao2019}.

\subsection{Miscellaneous applications}

The integral equations (\ref{GLLu}) appear in several other physical contexts. 
One of these is in the construction of solutions within little string theory, with $g(x)=x^n$ \cite[\eqn~(3.7)]{ling2006}
and $g(x)=1-\beta x^2$ \cite[\eqn~(C.5)]{ling2006}, where $\beta$ is a positive constant.
The same quadratic $g$ arises with multi\-component bosons \cite[\eqn~(47)]{Li2003}, 
in a super--Yang--Mills theory \cite[\eqn~(3.29)]{Lin2006}, \cite[\eqn~(3.8)]{vananders2007}, and in the zero-temperature limit of the Yang--Yang model \cite[p.~36, \eqn~(7.9)]{Korepin1993}. 
In the last two applications just mentioned \cite{Korepin1993,Lin2006}, the constant $\beta$ is chosen so that $u(1)=0$.
For a similar mathematical problem, with a rational $g$ and an application to spin chains, see \cite[\eqn~(4.21)]{Hao2019}.
 
Further applications include evaluating statistical properties of a two-dimensional lattice of elastic lines in a random medium \cite[\eqn~(53)]{Emig2001},
and
calculating the ground-state properties of the attractive two-component Hubbard model \cite[\eqn~(2.44)]{Marino2020}, in particular at half filling \cite[\eqn~(7)]{Takahashi1969}.

There are also applications in probability theory. 
For instance, in 1953, Reich \cite{Reich1953} showed that the Love--Lieb equation \LLE{-}{1} applies to a specific one-dimensional random walk with absorbing barriers: ``In addition to its theoretical interest, the random walk appears to provide a practical means for the calculation of the capacitance by a Monte Carlo technique''; for another application, see~\cite{Hwang2006}.

% *********************************************************************
\section{Solving Love--Lieb integral equations: basic theory}\label{techniques}
% **************************************************************

\subsection{Difficulties near the endpoints when $\alpha$ is small}\label{difficultSec}

Recall the generalized Love--Lieb equation (\ref{GLLu}), which we write as
\be
   u(x)\pm \int_{-1}^1 K(x-y)\,u(y)\,\rd y=g(x),\quad -1\leq x\leq 1, \label{GLLx}
\ee
where the kernel is given by
\be
   K(x)=\frac{\alpha}{\pi(\alpha^2+x^2)},\quad \alpha>0. \label{K-DEFN}
\ee
Equation (\ref{GLLx}) is classified as a Fredholm integral equation of the second kind with a continuous kernel. 
This is a textbook case \cite{Smithies, Cochran, Kress}:
standard theory applies and almost any sensible numerical method can be employed to solve it. 
However, difficulties are expected when $\alpha$ is small because $K(x)$ is
a well-known approximation to a Dirac delta:
\be
  \lim_{\alpha\to 0} \int_{-1}^1 K(x-y)\,f(y)\,\rd y=f(x),\quad -1<x<1.\label{KintLimit}
\ee
To see why these difficulties arise, start by considering the Love--Gaudin equation \LLE{+}{1}. 
Using (\ref{KintLimit}) in \LLE{+}{1} yields $u(x)\simeq \frac{1}{2}$ for $|x|<1$, whereas the integral equation itself gives
\begin{align*}
  u(1)&=1-\int_{-1}^1K(1-y)\,u(y)\,\rd y
  \simeq 1-\frac{1}{2}\int_{-1}^1K(1-y)\,\rd y \\
  &=1-\frac{1}{2\pi}\arctan\frac{2}{\alpha}
  =\frac{3}{4}+\frac{1}{2\pi}\arctan\frac{\alpha}{2},
\end{align*}
hence $u(1)\simeq\frac{3}{4}$ for $\alpha\ll 1$. Here, we have used
\be
  \int_{-1}^1K(x-y)\,\rd y=
  \frac{1}{\pi}\left[\arctan{\left(\frac{1-x}{\alpha}\right)}
  +\arctan{\left(\frac{1+x}{\alpha}\right)}\right]
  \equiv \Kc(x;\alpha),\label{Kintegral}
\ee
say.
This argument (which is taken from \cite[p.~26]{Phillips72}) shows that rapid variations in $u(x)$ near the endpoints are to be expected when $\alpha\ll 1$.

If we try to apply the same arguments to the Love--Lieb equation \LLE{-}{1}, we find that the two terms on the left-hand side cancel, so that we need a refined version of (\ref{KintLimit}). 
From \cite[\eqn~(21)]{Farina2010}, we have
\begin{equation}  \label{eq:hypersingular}
\int_{-1}^1 K(x-y)\,f(y)\,\rd y =f(x)
   + \frac{\alpha}{\pi} \fpint_{-1}^{1}\frac{f(y)}{(x-y)^2} \, \rd y
   +O(\alpha^2) \quad\mbox{as $\alpha\to 0$,}
\end{equation}
where $x$ is bounded away from $\pm 1$ and the cross on the integral denotes a hyper\-singular (finite-part) integral.

Of course, formulas such as (\ref{KintLimit}) and (\ref{eq:hypersingular}) do not take account of any possible variation of $f$ with $\alpha$.
Nevertheless, ignoring this complication does give some useful approximations to $u(x;\alpha)$.

We are going to use (\ref{eq:hypersingular}) to approximate the left-hand side of \LLE{-}{g}. Before doing that, we note that
\[
 \fpint_{-1}^{1}\frac{f(y)}{(x-y)^2}\,\rd y
 =-\frac{f(1)}{1-x}-\frac{f(-1)}{1+x} 
   +\cpvint_{-1}^1\frac{f'(y)}{y-x}\,\rd y,
\] 
where the integral on the right is a Cauchy principal value (CPV) integral.
If this formula is used in (\ref{eq:hypersingular}) {\it and it is assumed that $f(1)=f(-1)=0$\/}, we recover a formula used by Kac and Pollard \cite[Lemma~5.1]{kac1950} and by others. 
The connection with CPV integrals is attractive (because they are more familiar) but the condition on $f(\pm 1)$ is not satisfied by solutions
of \LLE{-}{g}, in general.

Formally, then, we obtain an approximation to \LLE{-}{g},
\[
-\frac{\alpha}{\pi}\fpint_{-1}^{1}\frac{u(y)}{(x-y)^2} \,\rd y=g(x),\quad -1<x<1.
\]
The general solution of this hypersingular integral equation is known \cite{PAMJIEA}. It consists of a particular solution (corresponding to the given $g$) together with the general solution of the homogeneous equation (put $g=0$), which is $(A+Bx)(1-x^2)^{-1/2}$, where $A$ and $B$ are arbitrary constants. As we do not want solutions that are unbounded at $x=\pm 1$, we take $A=B=0$.
In particular, for the Love--Lieb equation \LLE{-}{1}, we obtain $u(x)=\alpha^{-1}\sqrt{1-x^2}$.
This approximation to $u(x)$ is incorrect at the endpoints because it can be shown that $u(\pm 1)>1$; see (\ref{loweru}) below.

Similar approximations can be obtained for \LLE{\pm}{x}. For \LLE{+}{x}, we obtain $u\simeq x/2$ for $|x|<1$,
whereas for \LLE{-}{x} we obtain $u\simeq (2\alpha)^{-1} x\sqrt{1-x^2}$ (in agreement with (\ref{LxFORM}) below).

We shall return to analytical approximations of $u(x)\equiv u(x;\alpha)$ in Sec.~\ref{LLapproxSec}, where we also make comparisons with direct numerical solutions of the integral equation.

\subsection{Solvability, iteration and Liouville--Neumann expansions}

The general theory of Fredholm integral equations of the second kind such as (\ref{GLLx}) tells us to examine the homogeneous version of (\ref{GLLx});
following Hilbert, it is convenient to insert a parameter $\lambda$, giving
\be
   \psi(x)-\lambda \int_{-1}^1 K(x-y)\,\psi(y)\,\rd y=0,\quad -1\leq x\leq 1. \label{GLLz}
\ee
We are especially interested in $\lambda=1$ and $\lambda=-1$, because these special cases correspond to \LLE{-}{g} and \LLE{+}{g}, respectively.
As the kernel is symmetric, Hilbert--Schmidt theory 
\cite[section~7.2]{Smithies}, \cite[section~7.2]{Cochran}
states that there is at least one real value of $\lambda$ for which (\ref{GLLz}) has a non-trivial solution $\psi$.
Fortunately, such {\it characteristic values\/} include neither $\lambda=1$ nor $\lambda=-1$.
This was proved by Love \cite[Lemma~6]{Love1949}, using simple
iterated inequalities; actually, his argument shows that (\ref{GLLz})
has no non-trivial integrable solution for $|\lambda|<2$.
For more information on eigenvalues (reciprocals of characteristic values)
and eigenfunctions, see~\cite{baratchart2019}.

Returning to (\ref{GLLx}), we can apply the Fredholm Alternative
\cite[Theorem~3.6.1]{Smithies}, \cite[section~3.6]{Cochran}:
as the homogeneous version of (\ref{GLLx}) has no non-trivial solution,
the inhomogeneous integral equation (\ref{GLLx}) has exactly one solution
for any right-hand side function $g$. 
This result can be stated in terms of continuous or square-integrable functions. 
Love \cite[Lemma~7]{Love1949} gives this result for continuous solutions of \LLE{\pm}{1}.
Lieb and Liniger obtain the same result for \LLE{-}{g}, exploiting
the positivity of the kernel and the Liouville--Neumann expansion \cite[Appendix~B]{LiebLiniger1963}. 
This expansion arises when any Fredholm integral equation of the second kind is solved iteratively. 
To see this, consider
\be
   u(x)-\lambda\int_{-1}^1K(x-y)\,u(y)\,\rd y =g(x),\quad -1\leq x\leq 1.\label{LLLg}
\ee
The Liouville--Neumann expansion \cite[section~2.5]{Smithies}, \cite[section~3.1]{Cochran} for $u$ is
\be
  u(x)=g(x)+\sum_{n=1}^\infty \lambda^n \int_{-1}^1 K_n(x,y)\,g(y)\,\rd y, \label{LLLgLN}
\ee
where the iterated kernels $K_n$ are defined by $K_1=K$ and
\be
K_n(x,y)=\int_{-1}^1K_{n-1}(x,s)\,K(s,y)\,\rd y
=\int_{-1}^1K(x,s)\,K_{n-1}(s,y)\,\rd y \label{iteratedK}
\ee
for $n=2,3,\ldots$.
The expansion (\ref{LLLgLN}) can be recast as an iterative process; doing this for (\ref{LLLg}) gives
\be
  u_n(x)=g(x)+\lambda \int_{-1}^1K(x-y)\,u_{n-1}(y)\, \rd y,\quad n=1,2,\ldots, \label{LoveIterg}
\ee
starting with $u_0=g$, for example. 

The convergence of the series (\ref{LLLgLN}) was proved by Love \cite[Theorem~2]{Love1949} in the context of (\ref{LLeqn}), but his proof extends to (\ref{LLLg}); the condition for convergence is found to be
\[
   (2|\lambda|/\pi)\arctan{(1/\alpha)}<1,
\]
which is satisfied for $\lambda=\pm 1$ and $\alpha>0$, that is, for (\ref{GLLu}). Hafen \cite[p.~529]{Hafen10} wrote down the series (\ref{LLLgLN}) for (\ref{LLeqn}), but did not go further.
Love \cite{Love1949} also proved that the iterative process \ref{LoveIterg} is convergent; his proof shows that $u_n\to u$ as $n\to\infty$, where $u$ is the unique solution of (\ref{LLLg}).

It is worth noting that Love's convergence proofs rely on properties of the kernel $K$, (\ref{K-DEFN}). 
Textbook proofs for a general kernel $K$ require a condition such as $|\lambda|\| K\|<1$, where $\| K\|$ is the $L^2$-norm of $K$ \cite[Theorem~2.5.2]{Smithies}, but this condition is insufficient to establish convergence of (\ref{LLLgLN}) for all $\alpha>0$.

\subsection{Bounds}

Consider \LLE{-}{1}. Putting $g=1$ and $\lambda=1$ in
(\ref{LLLgLN}), and noting that all the iterated kernels are positive
(because $K(x)>0$), we infer that
\be
   u(x;\alpha)>1\quad \mbox{for $-1\leq x\leq 1$ and $\alpha>0$.}\label{loweru}
\ee
Other known bounds are on $|u|$. Thus Hutson \cite[p.~214]{Hutson1963}
proved the following bounds for \LLE{-}{g} but his proof extends to \LLE{+}{g}: assuming that $g(x)$ is a bounded continuous function, then
\begin{align}
    \sup_{-1\leq x\leq 1}|u(x;\alpha)|&\leq \sup_{-1\leq x\leq 1}\left\{ |g(x)| \left[1-\Kc(x;\alpha)\right]^{-1}\right\}, \label{H(2.6)}\\
    \sup_{-1\leq x\leq 1}|u(x;\alpha)|&\leq \sup_{-1\leq x\leq 1}\left\{ 
    \pi\, |g(x)| \,\frac{1-|x|+\alpha}{\alpha} \right\}, \label{H(2.7)}
\end{align}
with $\Kc$ defined by (\ref{Kintegral}).
These bounds hold for both \LLE{-}{g} and \LLE{+}{g}.
When $g=1$, the right-hand side of (\ref{H(2.6)}) reduces to $\pi/(2\arctan(\alpha))$, a bound found earlier by Reich \cite[p.~344]{Reich1953}.

\subsection{Maclaurin expansions}\label{MaclaurinSec}

We have seen that (\ref{LLLg}) is uniquely solvable for $u$, and so it is natural to ask if $u$ has a Maclaurin expansion.
For simplicity, consider (\ref{LLeqn}); take $g=1$ in (\ref{LLLg}).
In this case, $u$ is even, and its Maclaurin expansion takes the form
\be
  u(x;\alpha)=\sum_{n=0}^\infty c_n(\alpha)\,x^{2n},\quad |x|<\ell(\alpha),\label{uMaclaurin}
\ee
where $\ell(\alpha)$ is the radius of convergence and the coefficients $c_n(\alpha)$ are uniquely determined.
We are particularly interested in determining when  $\ell(\alpha)>1$ because then 
the solution of (\ref{LLeqn}) can be sought in the form (\ref{uMaclaurin}).

Start by expanding the kernel $K(x-y)$, (\ref{K-DEFN}), using the binomial theorem.
There are several options. 
One is to expand in powers of $(x-y)^2/\alpha^2$,
\be
  K(x-y)=\frac{1}{\pi\alpha}\sum_{n=0}^\infty \frac{(-1)^n}{\alpha^{2n}} (x-y)^{2n}. \label{KxyONE}
\ee
This converges for all $x$ with $|x|<1$ and for all $y$ with $|y|<1$ if $\alpha>2$.
The expansion (\ref{KxyONE}) can be seen as an expansion in inverse powers of $\alpha$.
For this reason, it has been used to obtain approximations for $u(x;\alpha)$
when $\alpha\gg 1$; see Sec.~\ref{largegapsec}.

For another option, use partial fractions and write $K$ as
\be
  K(x-y)=\frac{\ci}{2\pi}\left(
  \frac{1}{x-y+\ci\alpha}-
  \frac{1}{x-y-\ci\alpha}\right). \label{Kpartial}
\ee
Expanding in powers of $y/(x\pm\ci\alpha)$,
\be
K(x-y)=\frac{\ci}{2\pi}\sum_{n=0}^\infty y^n\left(
\frac{1}{(x+\ci\alpha)^{n+1}}
-\frac{1}{(x-\ci\alpha)^{n+1}}\right). \label{KxyTWO}
\ee
The series converges for $|y|<\sqrt{x^2+\alpha^2}$.
Therefore it converges for all $x$ with $|x|<1$ and for all $y$ with $|y|<1$ if $\alpha>1$. 
To simplify (\ref{KxyTWO}), define real quantities $X$ and $\varphi$ by $x\pm\ci\alpha=X\,\ce^{\pm\ci\varphi}$ so that $X=\sqrt{x^2+\alpha^2}$, $\cos\varphi=x/X$ and $\sin\varphi=\alpha/X$. 
Then (\ref{KxyTWO}) becomes
\begin{align}
K(x-y)&=\frac{\ci}{2\pi}\sum_{n=0}^\infty \frac{y^n}{X^{n+1}}\left(
\ce^{-\ci(n+1)\varphi}-\ce^{\ci(n+1)\varphi}\right)
=\frac{1}{\pi}\sum_{n=0}^\infty \frac{y^n}{X^{n+1}}\sin[(n+1)\varphi]\nonumber\\
&=\frac{\alpha}{\pi}\sum_{n=0}^\infty \frac{y^n}{X^{n+2}}\frac{\sin[(n+1)\varphi]}{\sin\varphi}
=\frac{\alpha}{\pi}\sum_{n=0}^\infty \frac{y^n}{X^{n+2}}\,U_n\!\left(\frac{x}{X}\right),\label{KxyUTWO}
\end{align}
where $U_n$ is a Chebyshev polynomial of the second kind
and we have used $x/X=\cos\varphi$.
This expansion (but without the identification of the Chebyshev polynomials)
was used by Wadati~\cite[\eqn~(2.2)]{Wadati2002}.

Interchanging $x$ and $y$ in (\ref{KxyUTWO}) gives
\be
  K(x-y)=\frac{\alpha}{\pi}\sum_{n=0}^\infty \frac{x^n}{Y^{n+2}}\,U_n\!\left(\frac{y}{Y}\right),\label{KxyUTHREE}
\ee
where $Y=\sqrt{y^2+\alpha^2}$.
As before, (\ref{KxyUTHREE}) converges for all $x$ with $|x|<1$ and for all $y$ with $|y|<1$ if $\alpha>1$.

If we substitute (\ref{KxyUTHREE}) in (\ref{LLLg})
(with $g=1$), assuming that $\alpha>1$, we find
\[
  u(x;\alpha)=1  +\frac{\lambda\alpha}{\pi} \sum_{n=0}^\infty x^n
  \int_{-1}^1 \frac{1}{Y^{n+2}}\,U_n\!\left(\frac{y}{Y}\right) u(y;\alpha)\,\rd y.
\]
The polynomial $U_n$ is even when $n$ is even and odd when $n$ is odd.
Then, as $u$ is even, we obtain
\[
  u(x;\alpha)=1  +\frac{\lambda\alpha}{\pi} \sum_{n=0}^\infty x^{2n}
  \int_{-1}^1 \frac{1}{Y^{2n+2}}\,U_{2n}\!\left(\frac{y}{Y}\right) u(y;\alpha)\,\rd y,
\]
giving the following formulas for the coefficients in (\ref{uMaclaurin}),
\begin{align}
  c_0(\alpha)&=1+\frac{\lambda\alpha}{\pi}\int_{-1}^1 \frac{u(y;\alpha)}{y^2+\alpha^2}\,\rd y,\label{c0INTREP}\\
  c_n(\alpha)&=\frac{\lambda\alpha}{\pi}\int_{-1}^1 \frac{u(y;\alpha)}{(y^2+\alpha^2)^{n+1}}\,
  U_{2n}\!\left(\frac{y}{\sqrt{y^2+\alpha^2}}\right)  \rd y,
  \quad n=1,2,\ldots.\label{cnINTREP}
\end{align}

The expansion (\ref{KxyUTHREE}) can also be used in (\ref{iteratedK}) to expand the iterated kernels for $\alpha>1$.

For $\alpha>1$, we may substitute (\ref{uMaclaurin}) in the right-hand sides of (\ref{c0INTREP}) and (\ref{cnINTREP}), leading to an infinite linear system for the unknown coefficients. 
Truncating this system would lead to a numerical method; see (\ref{c0Nformula}) and (\ref{cmNformula}).

%%%%%%%%%%%%%%%%%%%%%%%%%%%%%%%%%
\section{Solving Love--Lieb integral equations: numerical methods}\label{numerics}
%%%%%%%%%%%%%%%%%%%%%%%%%%%%%%%%%

In October 1925, there was a meeting of the American Mathematical Society  in Berkeley, California.
The title and complete abstract of a contributed paper are as follows~\cite{HB25abs}:
\smallskip
\begin{quotation}
\noindent
Professor Harry Bateman: \textit{Numerical solution of an integral equation}.

Hafen has shown that the distribution of electricity on the circular
plates of a parallel plate condenser can be found by solving a linear
integral equation of the second kind. In the present paper the author
solves this equation numerically by a method of least squares, and
discusses the method in a general way, remarking on the question of
convergence.
\end{quotation}
\smallskip
Evidently, the abstract refers to the integral equation (\ref{LLeqn}), and it may be the first ever attempt to tackle this problem numerically.
Bateman did not publish a full paper on what he did (although he published many papers on integral equations).
However, his abstract marks the beginning of almost a century of efforts to devise numerical methods to solve integral equations.
These efforts are the focus for this section in the context of Love--Lieb equations.

\subsection{Nystr\"om's method}
\label{Nystr}

Love's integral equation (\ref{LLeqn}) is simple and it is related to  physical problems, but its solution is not known in closed form.
These facts made it attractive to numerical analysts who were developing methods for solving integral equations using computers.
An early example is the paper by Fox and Goodwin \cite{FoxG1953}.
They used a method that is known nowadays as the {\it Nystr\"om method\/} \cite{Nystrom30}, \cite[Chapter~4]{Atkinson1997}:
approximate the integral using a quadrature rule and then collocate, leading to a linear algebraic system. Thus, using 
\be
  \int_{-1}^1 F(x)\,\rd x\simeq \sum_{j=1}^N w_j F(x_j),\quad
  -1\leq x_1<x_2<\cdots <x_N\leq 1, \label{quadrule}
\ee
with weights $w_i$ and nodes $x_i$, $i=1,2,\ldots, N$, (\ref{LLLg}) gives
\be
   u(x)\simeq g(x)+\lambda \sum_{j=1}^Nw_j u(x_j)\,K(x-x_j),\quad -1\leq x\leq 1. \label{NyInterp}
\ee
Collocation at $x=x_i$ then gives the $N\times N$ linear system
\[
  u_i -\lambda \sum_{j=1}^N w_j K(x_i-x_j) u_j = g(x_i),\quad i=1,2,\ldots, N,
\]
where $u_i\simeq u(x_i)$. Having computed $u_1, u_2,\ldots,u_N$, one can then use (\ref{NyInterp}) to approximate $u(x)$
(the right-hand side of (\ref{NyInterp}) is Nystr\"om's interpolation formula), or one can interpolate through
the $N$ computed values of $u(x)$ at the nodes.

For (\ref{LLeqn}) (with $g(x)=1$), we know that $u(x)$ is even, while for (\ref{LiebTwo}) (with $g(x)\equiv x$), $u(x)$ is odd. 
In both cases, the integral equation is easily converted into an equation that holds for $0\leq x\leq 1$, and then the size of the linear system can be halved for the same accuracy.

Simple choices for the quadrature rule (\ref{quadrule}) work well, at least when $\alpha$ is not too small (otherwise $u$ has sharp variations close to the endpoints, as mentioned before).
In their 1953 paper, Fox and Goodwin \cite{FoxG1953} used the repeated trapezoidal rule and gave numerical results when $\alpha=1$. 
A few years later, Cooke \cite{Cooke1958} tried the same method to solve the capacitor problem with a small gap:
\smallskip
\begin{quotation}
\noindent
Finding difficulty in using their method for $\alpha=0.1$, I approached Dr.~Fox, and he kindly consented to solve the problem for this $\alpha$.
Dr.\ J.~Blake carried out the work and found that it was necessary to divide the range of integration
into 50 parts and solve a system of 50 linear equations in 50 unknowns in order to obtain 4-figure accuracy!
Naturally use was made of high speed computing machinery.
\end{quotation}
\smallskip
Sixty years later, Prolhac \cite{Prolhac2017} used the same method together with Richardson extrapolation with respect to~$N$ (and modern computer hardware), thus obtaining solutions of high accuracy.

Other quadrature rules have been used for (\ref{LLeqn}); these include Simpson \cite{LiebLiniger1963},
Clenshaw--Curtis \cite{WintleK85} and Gauss--Legendre \cite{Paffuti17, LinShi18}.
For some comparisons (when $\alpha=1$), see~\cite{Boland72}.

Another option is to use a quadrature rule that has been designed to handle the kernel (\ref{K-DEFN}).
For applications of such product rules,
see \cite{Monegato88} and \cite{Fermo2020}; the paper \cite[Example~5.1]{Fermo2020} gives some numerical results for \LLE{+}{g}.

In the 1970s, software became available for solving integral equations such as \LLE{\pm}{g} using the Nystr\"om method.
Atkinson \cite{Atkinson1976} offered two FORTRAN programs, one using Simpson's rule (called IESIMP) and one using Gauss--Legendre quadrature (called IEGAUS). Both have been used to solve the Love--Lieb equation,
\LLE{-}{1}; IEGAUS was used in \cite{Dunjko2001} 
and IESIMP in \cite{Girardeau2003}.
More recently, Atkinson and Shampine \cite{Atkinson2008} updated and extended IESIMP into a {\sc Matlab} program called \texttt{Fie}. It has been used in \cite{Shamailov2016} for \LLE{-}{g},
and we have used it for \LLE{\pm}{1} and for \LLE{-}{x}; see Sec.~\ref{sanrSec}.

The kernel $K(x-y)$ has singularities at $y=x\pm \ci\alpha$. 
These two points are not in the range of integration, but they become closer as $\alpha$ becomes smaller. 
Therefore, it can be helpful to use a regularization \cite{WintleK85, Paffuti17}, writing (\ref{LLLg}) as
\[
  u(x)\left(1-\lambda\int_{-1}^1 K(x-y)\,\rd y\right) -\lambda \int_{-1}^1 K(x-y) \left\{u(y)-u(x)\right\}\rd y=g(x),\quad |x|<1;
\]
for the first integral, see (\ref{Kintegral}).

%%%%%%%%%%%%%%%%%%%%%%%%%%%%%%%%%
\subsection{Iterative methods}

The iterative process (\ref{LoveIterg}) can be used to solve (\ref{LLeqn}) numerically, using a quadrature rule (\ref{quadrule}) to approximate the integrals.
For a straightforward implementation, see~\cite{Bartlett85}.

Love \cite{Love1990} returned to (\ref{LoveIterg}) for (\ref{LLeqn}) but he introduced an extra step.
He took the numerical results obtained by Fox and Goodwin \cite{FoxG1953}
for $u$ when $\alpha=1$, interpolated using an even polynomial of degree $8$, and then took this as his initial guess for $u_0$. 
Richardson \cite{Richardson2004} used the same method, starting with $u_0=1$ and interpolating at each step of the iterative process.

%%%%%%%%%%%%%%%%%%%%%%%%%%%%%%%%%%%%%%%%%%%%%%%%%
\subsection{Expansion methods}\label{ExpMethSec}

Another class of numerical methods starts by expanding $u$ using a set of basis functions,
\be
   u(x)\simeq \sum_{n=0}^N c_n \, \Phi_n(x), \quad -1<x<1,\label{uNEXPAND}
\ee
with coefficients $c_n$ and basis functions $\Phi_n(x)$, $n=0,1,2,\ldots$.
Substitution in (\ref{LLLg}) gives
\be
\sum_{n=0}^Nc_n \left( \Phi_n(x)-\lambda \int_{-1}^1 K(x-y)\,\Phi_n(y)\,\rd y
\right)\simeq g(x),\quad -1<x<1.  \label{GLLexp}
\ee
To proceed, one needs to (i) choose the functions $\Phi_n$ and then (ii) choose a way to determine $c_n$.
Let us start with (ii). One possibility is to use a Galerkin method: multiply (\ref{GLLexp}) by $\Phi_m(x)$ (with $m=0,1,2,\ldots,N$) and then integrate over the interval $-1<x<1$, giving a square linear algebraic system for the coefficients $c_n$.

A simpler choice is collocation: evaluate (\ref{GLLexp}) at $M+1$ points in the interval $-1\leq x\leq 1$ (with $M\geq N$), giving $M+1$ equations in the $N+1$ unknowns, $c_0, c_1,\ldots, c_N$;
if $M=N$, we obtain a square system, whereas if $M>N$, the system is over\-determined and we may use least-squares to obtain an approximate solution (perhaps as Bateman did in 1925).

Let us now consider (i), the choice of the functions $\Phi_n$.
There are many options, such as trigonometric functions, monomials or orthogonal polynomials; we shall discuss each of these below.

For simplicity, let us assume that $g$ is even so that $u$ is also even. Then, one natural choice is to try
\[
   \Phi_n(x)=\cos{(n\pi x)},
\]
since these functions are even and orthogonal over the interval $[-1,1]$. 
A Galerkin method (multiply (\ref{GLLexp}) by $\cos{(m\pi x)}$ and integrate over $-1<x<1$) then leads to an algebraic system for the coefficients $c_n$. 
This approach has been employed in \cite{Carlson1994} and \cite{norgren2009}; in the latter paper, the relevant double integrals are evaluated analytically in terms of sine and cosine integrals.

Chebyshev polynomials were used by Elliott \cite{Elliott63}, with the choice
\[
  \Phi_n(x)=T_{2n}(x)
\]
for  Love's equation.  Piessens and Branders \cite{PiessensB76}
showed how to compute the integrals
\be
   I_n(x)=\int_{-1}^1 K(x-y)\,T_n(y)\,\rd y \label{InKTn}
\ee
recursively; see also \cite{Ristivojevic2019}.
For the use of certain close relatives of Chebyshev polynomials, see~\cite{Milovanovic2013, Vellucci16}. 

Infinite series of Legendre polynomials $P_n$ were used by Love \cite[\eqn~(16)]{Love1949},
\be
  u(x)=\sum_{n=0}^\infty c_n \, P_{2n}(x), \label{ERL(16)}
\ee
but they do not play a significant role in his analysis:
``We need not consider in what sense the series (\ref{ERL(16)}) is to be understood if it is not
convergent; for the results of the present formal work are rigorously verified later'' \cite[p.~436]{Love1949}.
The infinite series (\ref{ERL(16)}) has been used more recently
\cite{Reichert2020, Lang2017, Ristivojevic2014}. Numerical results obtained with the choice $\Phi_n(x)=P_{2n}(x)$ have also been reported \cite{Zhou2012}.

Instead of  orthogonal polynomials, another option is to use simple monomials,
\be
   \Phi_n(x)=x^{2n},\label{monomials}
\ee
giving a polynomial approximation to $u$ by truncating the Maclaurin series thereby obtained.
The relevant integrals (replace $T_n(y)$ by $y^{2n}$ in (\ref{InKTn})) can be
evaluated explicitly \cite[\eqn~(64)]{Fabrikant87}, \cite[Appendix~D]{Lang2017}. 
This method has been implemented \cite{Olshanii2017}, \cite[section~B.3]{Lang2018}. 
The integrals mentioned above can be used to construct $g(x)$ so that $u(x)=x^{2n}$ solves \LLE{\pm}{g}.
Similar calculations can be made for other simple choices for~$u$.

Another way of using monomials (\ref{monomials}), already mentioned in Sec.~\ref{MaclaurinSec}, combines (\ref{uNEXPAND}) with (\ref{c0INTREP}) and (\ref{cnINTREP}) to give
\begin{align}
     c_0&=1+\frac{\lambda\alpha}{\pi}\sum_{n=0}^Nc_n\int_{-1}^1 \frac{y^{2n}}{y^2+\alpha^2}\,\rd y,\label{c0Nformula}\\
  c_m&=\frac{\lambda\alpha}{\pi}\sum_{n=0}^Nc_n\int_{-1}^1 \frac{y^{2n}}{(y^2+\alpha^2)^{m+1}}\,
  U_{2m}\!\left(\frac{y}{\sqrt{y^2+\alpha^2}}\right)  \rd y,
  \quad m=1,2,\ldots,N.\label{cmNformula}
\end{align}
Future implementations of this approach are to be expected.

Splines were first used for Love's integral equation (\ref{LLeqn}) by Phillips \cite{Phillips72}. 
They have been used more recently in \cite{barrera2018} for \LLE{-}{g} and in \cite{barrera2020} for~\LLE{+}{1}.
Although splines can be attractive in other contexts, for Love--Lieb integral equations, it is unclear that they are competitive with other 
numerical methods such as the Nystr\"om method.

\subsection{Element methods}

Partition the interval $[-1,1]$ into $N$ subintervals using
$-1=x_0<x_1<\cdots <x_{N-1}<x_N=1$, so that $x_{n-1}<x<x_n$ is the $n$th subinterval (element), $E_n$, $n=1,2,\ldots, N$. 
Then we can write (\ref{LLLg}), exactly, as
\be
  u(x)-\lambda\sum_{j=1}^N \int_{E_j} K(x-y)\,u(y)\,\rd y=g(x),\quad -1<x<1. \label{GLLxE}
\ee
If we evaluate (\ref{GLLxE}) for $x\in E_i$, $i=1,2,\ldots, N$, we obtain a coupled system of integral equations for $u$ on each element.

Approximating $u(x)\simeq u_i$, a constant, for $x\in E_i$, and then collocating (\ref{GLLxE}) at $x=\frac{1}{2}(x_{i-1}+x_i)$, the midpoint of the $i$th element, the result is a linear system for the numbers $u_i$, $i=1,2,\ldots,N$; the integrals encountered are similar to (\ref{Kintegral}).
This method was used by Wintle~\cite{Wintle86}.
Pastore \cite{Pastore2011} has used a more elaborate version of this method for~\LLE{+}{1}.

%%%%%%%%%%%%%%%%%%%%%%%%%%%%%%%%%%
\section{Solving Love--Lieb integral equations: approximations}\label{LLapproxSec}
%%%%%%%%%%%%%%%%%%%%%%%%%%%%%%%%%%

In this section, we discuss approximations to $u(x;\alpha)$ for small or large values of the parameter $\alpha$.
Much of the physical literature is aimed at approximating integrated quantities, such as the capacitance of the circular plate capacitor (\ref{CCap}) or the ground-state energy of the Lieb--Liniger model (\ref{energyLL}). 
Here, we limit ourselves to approximating $u(x;\alpha)$ itself.

\subsection{Approximations for $\alpha \gg 1$: large gaps, strong coupling}\label{largegapsec}

Consider the Love--Lieb equation, \LLE{-}{1}, with $\alpha\gg 1$. The solution is even, so look for the solution in the form~\cite{Ristivojevic2014}
\be
u(x;\alpha)\simeq
\sum_{n=0}^{2M+2}\frac{1}{\alpha^n} \sum_{m=0}^{M}c_{mn}\,x^{2m}, \label{Rcmn}
\ee
with coefficients $c_{mn}$ that do not depend on $M$.
The first terms beyond the Tonks--Girardeau limit ($c_{00}=1$) \cite{Girardeau1960} were explicitly obtained in \cite{Wadati2002, Zvonarev2005, rao2005}.
An algorithmic method to find $c_{mn}$ in a systematic way was developed by Ristivojevic \cite{Ristivojevic2014}. 
(In fact, he starts with (\ref{ERL(16)}).) 
His method steps forward in $M$, results for lower values of $M$ being used for higher values of $M$; systems of linear algebraic equations have to be solved at each step.
This algorithm was implemented up to $M=3$ in \cite{Ristivojevic2014}, recovering the results of \cite{rao2005}.
It was then implemented up to $M=8$ in \cite{Lang2017}, where the main properties of the algorithm have been discussed. 
The coefficients $c_{mn}$ are found to be polynomials in $1/\pi$ with rational coefficients; for an example, see (\ref{LLuEXP}) below.

A similar procedure yields the large-$\alpha$ approximation to the solution of the second Lieb equation \LLE{-}{x} (an odd function of $x$)
\cite{Ristivojevic2014},  \cite[Appendix~F]{Lang2017}. 

The method developed by Ristivojevic is quite complicated.
Let us now outline a simpler method, which is applicable to \LLE{\pm}{g}.
It starts by noting that if $\alpha>2$, then the kernel $K(x-y)$ can be expanded using the binomial theorem, (\ref{KxyONE}). 
If $g$ does not depend on $\alpha$,
the integral equation for $u$, (\ref{LLLg}), can then be used to show that $u(x;\alpha)$ has an expansion in powers of $\alpha^{-1}$:
\be
  u(x;\alpha)=g(x)+\frac{\lambda}{\pi}\sum_{n=0}^\infty \frac{(-1)^n}{\alpha^{2n+1}}\int_{-1}^1 (x-y)^{2n}\,u(y;\alpha)\,\rd y,
  \quad -1<x<1,\quad \alpha>2.\label{IEQx}
\ee
To solve this equation, put
\begin{align}
  u(x;\alpha)&=\sum_{n=0}^\infty \frac{u_n(x)}{\alpha^n}\label{uEXPalpha}\\
  &=\sum_{q=0}^\infty \left(\frac{u_{2q}(x)}{\alpha^{2q}}+\frac{u_{2q+1}(x)}{\alpha^{2q+1}}\right) \label{uEXPa}\\
  &=u_0(x)+\sum_{m=0}^\infty \left( \frac{ u_{2m+1}(x)}{\alpha^{2m+1}} +\frac{u_{2m+2}(x)}{\alpha^{2m+2}}\right)
  \label{uEXPb}
\end{align}
and substitute (\ref{uEXPa}) in the right-hand side of (\ref{IEQx}). 
We obtain
\[
u(x;\alpha)=g(x)+\frac{\lambda}{\pi}\sum_{q=0}^\infty\sum_{n=0}^\infty \frac{(-1)^n}{\alpha^{2n+2q+1}}
\int_{-1}^1(x-y)^{2n}\left(u_{2q}(y)+\frac{u_{2q+1}(y)}{\alpha}
\right)\rd y.
\]
In the double sum, put $n+q=m$ and then change the order of summation. 
Using (\ref{uEXPb}) on the left-hand side, we obtain
\begin{align*}
&
u_0(x)+\sum_{m=0}^\infty \left( \frac{ u_{2m+1}(x)}{\alpha^{2m+1}} +\frac{u_{2m+2}(x)}{\alpha^{2m+2}}\right)\\
&\qquad\mbox{}
=g(x)+
\frac{\lambda}{\pi}\sum_{m=0}^\infty\sum_{q=0}^m  \frac{(-1)^{m+q}}{\alpha^{2m+1}}\int_{-1}^1(x-y)^{2m-2q}\,u_{2q}(y)\,\rd y\\
&\qquad\qquad\mbox{}+\frac{\lambda}{\pi}\sum_{m=0}^\infty\sum_{q=0}^m  \frac{(-1)^{m+q}}{\alpha^{2m+2}}\int_{-1}^1(x-y)^{2m-2q}\,u_{2q+1}(y)\,\rd y.
\end{align*}
Matching powers of $\alpha$ then gives $u_0(x)=g(x)$,
\begin{align}
u_{2m+1}(x)&= \frac{\lambda}{\pi}\sum_{q=0}^m  (-1)^{m+q}\int_{-1}^1(x-y)^{2m-2q}\,u_{2q}(y)\,\rd y,  \label{uODD}\\
u_{2m+2}(x)&=\frac{\lambda}{\pi}\sum_{q=0}^m (-1)^{m+q}\int_{-1}^1(x-y)^{2m-2q}\,u_{2q+1}(y)\,\rd y,\label{uEVEN}
\end{align}
for $m=0,1,2,\ldots$.
These formulas (which appear to be new) show that both $u_{2m+1}(x)$ and $u_{2m+2}(x)$ are polynomials in $x$ of degree $2m$.
It turns out that the explicit expressions for $u_n$ simplify considerably when $g$ is even or odd.
The method extends readily to the case where $g$ depends on $\alpha$, provided $g(x;\alpha)$ can itself be expanded in inverse powers of~$\alpha$.

Let us calculate the first few terms. With the notation
$g_n=\int_{-1}^1 y^n\,g(y)\,\rd y$
and $\chi=\lambda/\pi$, we find
\begin{align*}
&
   u_0(x)=g(x), \quad u_1=\chi g_0, \quad u_2=2\chi^2g_0,\\
& u_3(x)=\chi(4\chi^2g_0-g_2)+2x\chi g_1-x^2\chi g_0, \\
&  u_4(x)=2\chi^2\{(4\chi^2-2/3)g_0-g_2\}-2x^2\chi^2g_0. 
\end{align*}
As an example, take $g(x)=1$ and $\lambda=1$, giving the Love--Lieb equation \LLE{-}{1}. Then
\be
  u_0=1,\quad u_1=\frac{2}{\pi},\quad
  u_2=\frac{4}{\pi^2},\quad u_3(x)=\frac{8}{\pi^3}-\frac{2}{3\pi}-\frac{2x^2}{\pi},\quad
  u_4(x)=\frac{16}{\pi^4}-\frac{4}{\pi^2}-\frac{4x^2}{\pi^2}. \label{LLuEXP}
\ee
These agree with \cite[\eqn~(2.3.53)]{Zvonarev2005}
and \cite[\eqn~(15)]{Ristivojevic2014}. Earlier, Wadati \cite{Wadati2002}
found an approximation in the form $u(x;\alpha)\simeq  a_0(\alpha)+a_2(\alpha)x^2$; approximating his solutions
for $a_0$ and $a_2$ \cite[\eqn~(3.3)]{Wadati2002} in powers of $\alpha^{-1}$ gives precise agreement with~(\ref{LLuEXP}).

For Gaudin's equation, \LLE{+}{1}, take $g(x)=1$ and $\lambda=-1$. This change to the sign of $\lambda$ has no effect on $u_0$, $u_2$ and $u_4$, but it changes the sign of $u_1$ and $u_3$.

Next, consider the second Lieb equation, \LLE{-}{x}: put $g(x)=x$ and $\lambda=1$. As $g_0=g_2=0$ and $g_1=\frac{2}{3}$,
we obtain
\be
  u_0(x)=x,\quad u_1=u_2=0,\quad 
  u_3(x)=\frac{4x}{3\pi},\quad u_4=0.  \label{L2uEXP}
%%   \quad u_5(x)=-\frac{8x}{\pi}\left( \frac{x^2}{3}+\frac{1}{5}\right).
\ee
These agree with  \cite[\eqn~(16)]{Ristivojevic2014}.

The polynomials $u_n(x)$ in (\ref{uEXPalpha}) are given recursively by (\ref{uODD}) and (\ref{uEVEN}).
As an alternative, we could seek the coefficients in the polynomials by direct substitution in the integral equation, much as was done with (\ref{Rcmn}).
But now, if we consider the Love--Lieb equation (\ref{LLeqn}), whose solution is even, and truncate (\ref{uEXPalpha}), we find that (\ref{Rcmn}) should be replaced by
\be
\label{dev}
u(x;\alpha)\simeq
\sum_{n=0}^{2M+2}\frac{1}{\alpha^n} \sum_{m=0}^{p(n)}c_{mn}\,x^{2m},
\ee
where the upper limit in the inner sum is $p$ when $n=2p+1$ or $2p+2$;
the additional terms in (\ref{Rcmn}) must all be zero.

\subsection{Approximations for  $\alpha\ll 1$: small gaps, weak coupling}\label{SmallalphaSec}

This limit is more difficult to handle because of the near-singularity of the kernel.
We have already seen that for \LLE{+}{1}, $u(x)\simeq \frac{1}{2}$ for $-1<x<1$ but $u(\pm 1)\simeq\frac{3}{4}$.
For \LLE{-}{1}, $u(x)\simeq \alpha^{-1}\sqrt{1-x^2}$  for $-1<x<1$ but $u(\pm 1)>1$; see (\ref{loweru}).
The errors in the approximations for $|x|<1$ close to the endpoints at $x=\pm 1$ suggest strongly using matched asymptotic expansions, and that is what we find in much of the literature.

Let us start with \LLE{-}{1}.
The leading (outer) approximation, away from the endpoints, is
\[
   u(x;\alpha)\simeq \alpha^{-1}\sqrt{1-x^2},\quad -1<x<1.
\]
This ``semi-circular law'' can be found in the 1963 papers by Lieb and Liniger \cite{LiebLiniger1963} (where it is a ``guess'') and Hutson \cite{Hutson1963} (where it is justified). 
Other proofs were proposed later \cite{Gaudin1971}, \cite[\eqn~(4.4)]{Wadati2002}, \cite{Olshanii2017}.
A more accurate (outer) approximation is
\be
 u(x;\alpha)\simeq
  \frac{\sqrt{1-x^2}}{\alpha}+\frac{1}{2\pi\sqrt{1-x^2}}\left[x\log\left(\frac{1-x}{1+x}\right)+\log\left(\frac{16\pi}{\alpha}\right)+1\right],
  \quad -1<x<1.\label{H(4.7)}
\ee
For derivations, see \cite[\eqn~(4.7)]{Hutson1963} and \cite[\eqn~(1.13)]{Popov1977}.

The two-term approximation (\ref{H(4.7)}) is integrable for $|x|<1$, and it can be used to obtain approximations for the capacitance (\ref{CCap}) or the ground-state energy (\ref{energyLL}). 
Higher-order terms can be added to (\ref{H(4.7)}) but they are not integrable and so the associated inner approximations (near the endpoints) are required~\cite{Reichert2020}.

For \LLE{+}{1}, the two-term approximation is
\be
  u(x;\alpha)\simeq \frac{1}{2}+\frac{\alpha}{2\pi(1-x^2)},
  \quad -1<x<1.\label{AL(3.7)}
\ee
For derivations, see \cite[\eqn~(15.23)]{Gaudin1968}, \cite[\eqn~(3.7)]{Atkinson1983} and \cite[\eqn~(3.82)]{Iida2007}.
Evidently, the approximation (\ref{AL(3.7)}) is not integrable for $|x|<1$, so that inner approximations are needed. 
An ansatz for the solution of \LLE{+}{1} was given in \cite{Reis2019bis}. Some of its coefficients can be fixed by a rather complicated procedure that matches inner and outer approximations. In particular, an outer approximation for $u$ is obtained \cite[\eqn~(3.29)]{Reis2019bis} that contains more terms than (\ref{AL(3.7)}).

For \LLE{-}{x}, the second Lieb equation, we are aware of two attempts, both leading to approximations in the form
\begin{align}
\label{LxFORM}
&  u(x;\alpha) \simeq  \frac{x}{2\alpha}\sqrt{1-x^2}\\
  &\ \mbox{}
   +\frac{\Ac(x)}{4\pi\sqrt{1-x^2}}
   \left[1+\log\left(\frac{16\pi}{\alpha}\right)\right]
   -\frac{\Bc(x)}{4\pi}\log{\left(\frac{1+x}{2}\right)}
  +\frac{\Bc(-x)}{4\pi}\log{\left(\frac{1-x}{2}\right)}\nonumber
\end{align}
for $-1<x<1$.
We know that $u(x;\alpha)$ is an odd function of $x$,
so that $\Ac(x)$ must also be odd, but $\Bc(x)$ is unrestricted.
Hutson \cite{Hutson64}, extending his analysis for \LLE{-}{1} \cite{Hutson1963}, obtained (\ref{LxFORM}) with
\be
  \Ac(x)=\sqrt{\frac{1+x}{2}}-\sqrt{\frac{1-x}{2}}
  \quad\mbox{and}\quad
  \Bc(x)=\frac{1}{\sqrt{2(1+x)}}.\label{Hut2Lieb}
\ee
More recently, Reichert et al.~\cite[\eqn~(S8)]{Reichert2019}
obtained (\ref{LxFORM}) but with different expressions for $\Ac$ and $\Bc$,
\be
  \Ac(x)=x \quad\mbox{and}\quad
  \Bc(x)=\frac{2x^2-1}{\sqrt{1-x^2}}. \label{R(S8)}
\ee
These authors used Popov's method \cite{Popov1977}, which starts with an assumed ansatz; 
implicit in their choice is that $\Bc(x)$ is an even function of $x$, which does not accord with Hutson's result~(\ref{Hut2Lieb}).

\subsection{Numerical results}\label{sanrSec}

We have given small-$\alpha$ approximations for the solutions of \LLE{\pm}{1}. Here, we compare them with direct numerical solutions of the integral equations, using Nystr\"om's method (\ref{Nystr}),
and the \textsc{Matlab} program \texttt{Fie} \cite{Atkinson2008}.
(The code was retrieved from \cite{atkinson2020}.)

\begin{figure}
\centering
\includegraphics{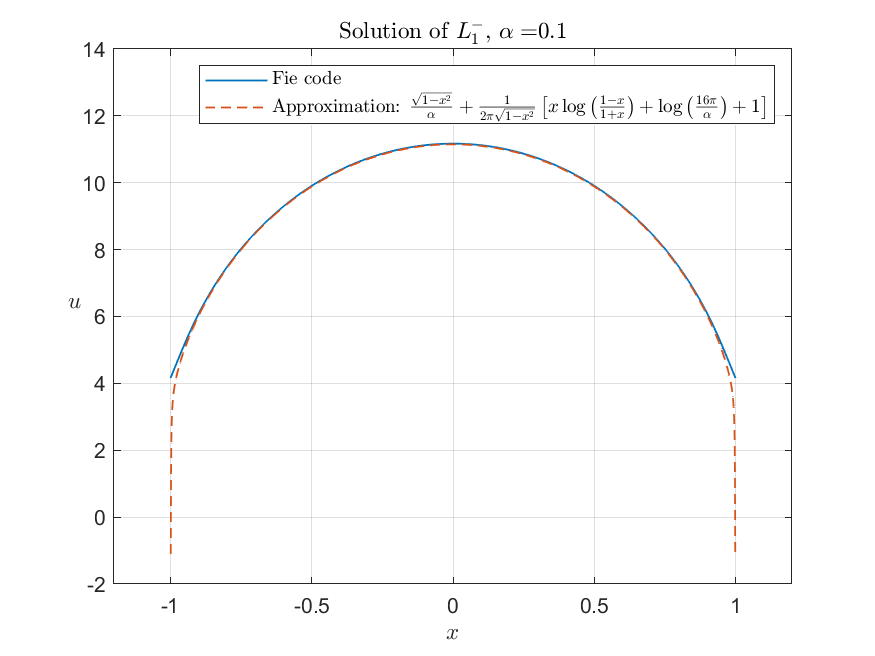}
\caption{The solution of the Love--Lieb equation {\rm \LLE{-}{1}} for $\alpha = 0.1$. 
The solid line represents the results obtained by the {\tt Fie} \textsc{Matlab} code and the dashed line shows the approximation given by (\ref{H(4.7)}).}
\label{fig:lovealpha01}
\end{figure}

\begin{figure}
\centering
\includegraphics[scale=1.0]{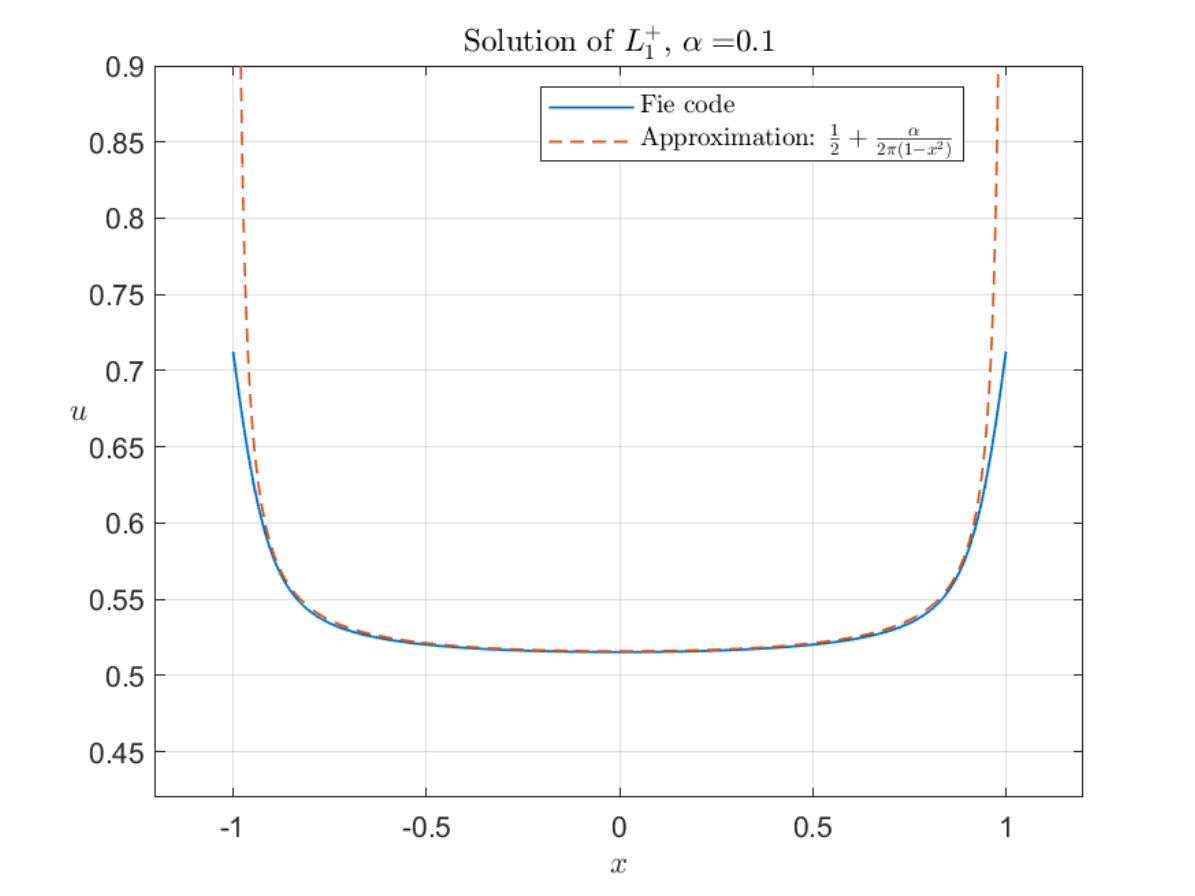}
\caption{The solution of the Gaudin equation {\rm \LLE{+}{1}}  for $\alpha = 0.1$. 
The solid line represents the results obtained by the {\tt Fie} \textsc{Matlab} code and the dashed line shows the approximation given by (\ref{AL(3.7)})}
\label{fig:gaudinalpha01}
\end{figure}

In Fig.~\ref{fig:lovealpha01}, the numerical solution of the Love--Lieb equation \LLE{-}{1} is plotted together with the approximation (\ref{H(4.7)}) for $\alpha=0.1$. The number of nodes used in Simpson's rule is 128.
Similarly, in Fig.~\ref{fig:gaudinalpha01}, the numerical solution of
Gaudin's equation \LLE{+}{1} is plotted together with the approximation (\ref{AL(3.7)}) for $\alpha=0.1$. The number of nodes used in the quadrature is 64. In both \texttt{Fie} results shown in these figures, the absolute and relative error tolerances are $10^{-6}$ and $10^{-3}$, respectively.

\begin{figure}
\centering
\includegraphics[scale=1.0]{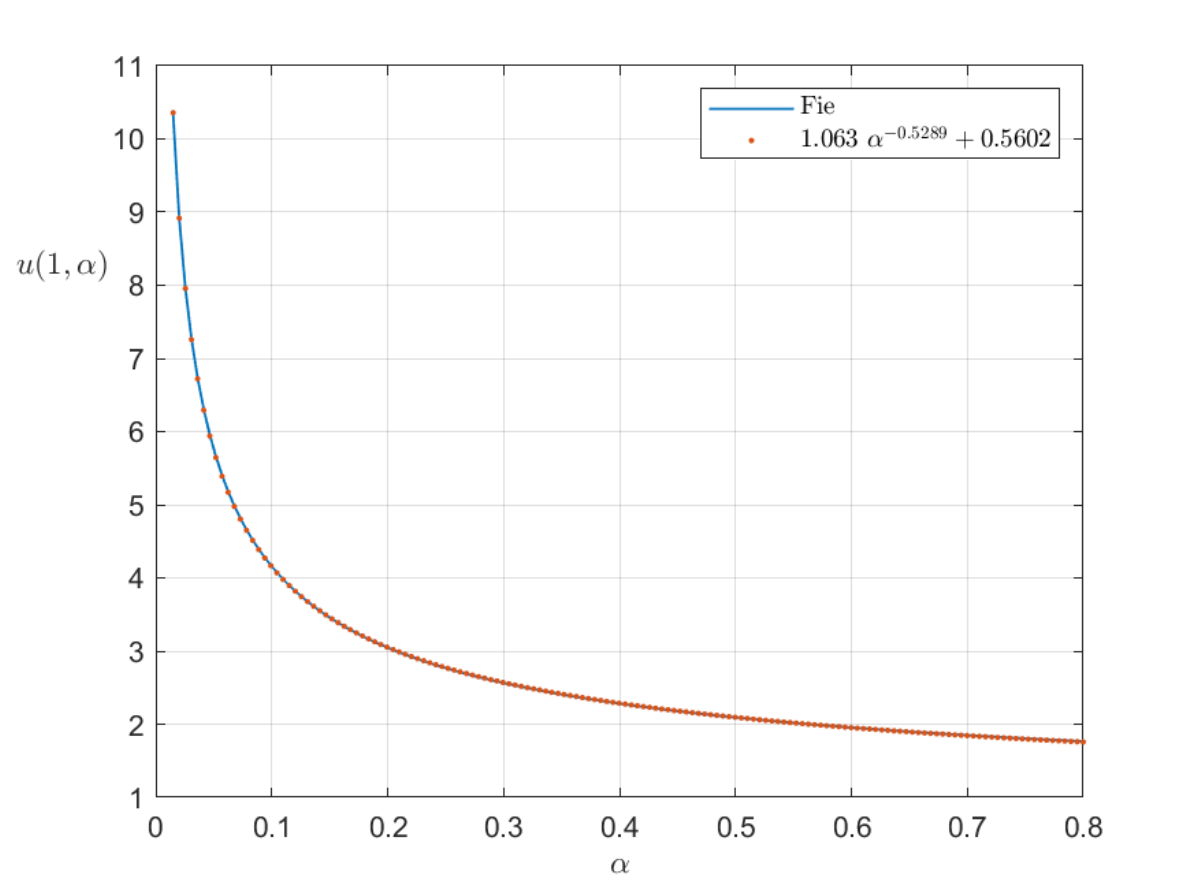}
\caption{The solution of {\rm \LLE{-}{1}} at $x = 1$ as a function of $\alpha$ in the domain $[0.05, 0.8]$. 
The solid line represents the results obtained by the {\tt Fie} \textsc{Matlab} code and the dashed line shows the fitting curve $1.063\alpha^{-0.5289} + 0.5602$, with a root mean square error of $0.0059$.}
\label{fig:lovefit}
\end{figure}

We see from the numerical results that the endpoint behaviour, at $x=\pm 1$, is not well captured by the outer approximations, as expected.
See also the remarks in Sec.~\ref{difficultSec}. We are not aware of any analytical approximations for $u(1;\alpha)$ as a function of $\alpha$.
Instead, as motivation for further study, we have fit 
%A starting point for studying the endpoints behaviour and getting an inner approximation could be fitting 
a curve by least squares to the numerical solutions by $\texttt{Fie}$. 
We have done that for \LLE{-}{1}; the result is shown in Fig.~\ref{fig:lovefit}. 
Among several approximating functions in the \textsc{Matlab} Curve Fitting Toolbox{\footnotesize \texttrademark}, the two-term power curve provided the smallest root mean square error.

\begin{figure}[!]
\centering
\includegraphics[scale=0.82]{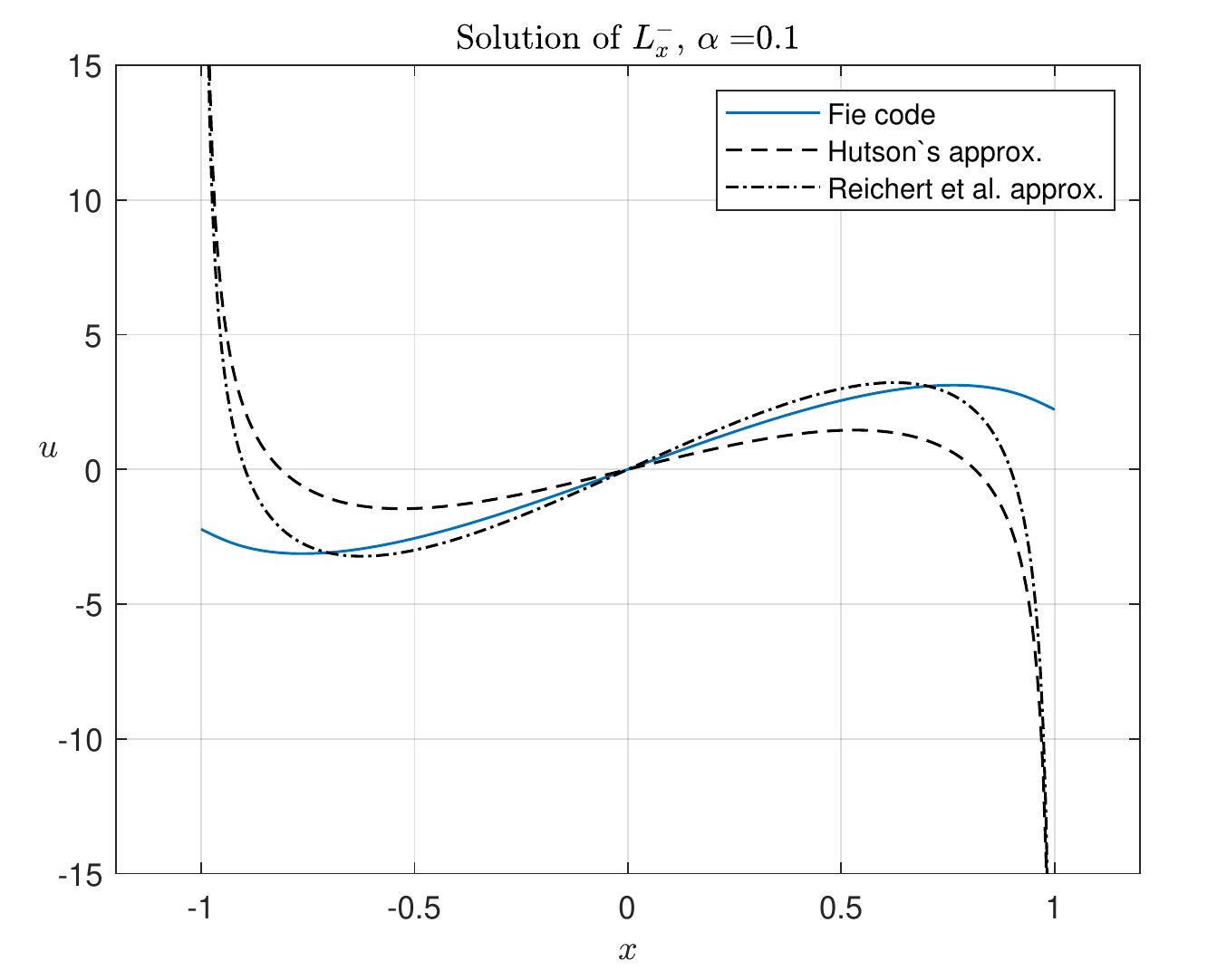}
\caption{The solution of {\rm \LLE{-}{x}} at $\alpha=0.1$ as a function of $x$. The solid line represents the results obtained by the {\tt Fie} \textsc{Matlab} code, the dashed line shows the approximation given by (\ref{Hut2Lieb}) and the dashed-dotted line (\ref{R(S8)})}
\label{fig:secondlieb}
\end{figure}

Numerical solutions of the second Lieb equation \LLE{-}{x} are plotted in Fig.~\ref{fig:secondlieb}, together with the approximations given by (\ref{Hut2Lieb}) and (\ref{R(S8)}).
It appears that (\ref{R(S8)}) is a better approximation, although further work is needed to resolve the discrepancies.

\section{Concluding remarks}
\label{cclSec}

We have reviewed 110 years of the literature on a very simple integral equation usually associated with the names of E.~R.~Love and E.~H.~Lieb. 
This equation has many applications in classical and quantum physics, as do some of its generalizations.
Despite its simplicity, no closed-form solution is known.
The study of this equation has inspired the developments of numerical and analytical methods, some of which exploit the size of $\alpha$, which is the only parameter appearing in the integral equation.
Further developments can be expected.

\section{Acknowledgments}

G.L.\ thanks Vladimir Korepin for drawing his attention to \cite{Hao2019} and the appearance of a generalized Love--Lieb equation in the spin chain context.
He thanks Etienne Granet for \cite{Griffiths1964},
which drew our attention to \cite{Hulthen38},
and Vanja Dunjko for enlightening discussions on \cite{Atkinson1976} and~\cite{Dunjko2001}.

\appendix
\section{Love--Lieb integral equations on an infinite interval}
\label{loveinftySec}

Consider the  integral equations
\be \label{eq:loveInfty}
   u(x) \pm  \int_{-\infty}^\infty K(x-y)\,u(y)\,\rd y  = g(x), \quad -\infty<x<\infty,
  \tag{\mbox{${\rm I}_g^\pm$}}
\ee
where
$K(x)=(\alpha/\pi)(x^2+\alpha^2)^{-1}$ and $\alpha>0$.
Formally, at least, we can solve (\ref{eq:loveInfty}) using Fourier transforms;
see, for example, \cite[section~11.1]{ECT}, \cite[section~8.5]{MF}, \cite[section~18.1]{Cochran}. Define
\[
  \ut(k)=\int_{-\infty}^\infty u(x)\,\ce^{\ci kx}\,\rd x.
\]
Then, using the convolution theorem, the Fourier transform of \ref{eq:loveInfty} is
\[
   \Delta_\pm(k)\,\ut(k)=\gt(k)
\]
with 
\[
  \Delta_\pm(k)=1\pm\Kt(k)
  \quad\mbox{and}\quad \Kt(k)=\ce^{-\alpha|k|}.
\]
Hence
\be
  \ut(k)=\frac{\gt(k)}{1\pm \ce^{-\alpha|k|}}. \label{utgt}
\ee  
Rearranging (as the denominator $\mbox{}\to 1$ as $|k|\to \infty$),
\be  
  \ut(k)=\gt(k)\mp\Mt_\pm(k)\,\gt(k)
  \quad\mbox{with}\quad \Mt_\pm(k)=\frac{\ce^{-\alpha|k|}}{1\pm\ce^{-\alpha|k|}}.
  \label{utgM}
\ee
Inverting, using the convolution theorem again,
\be
  u(x)=g(x)\mp\int_{-\infty}^\infty M_\pm(x-y)\,g(y)\,\rd y. \label{ugMCT}
\ee
At this stage, all these calculations are formal, of course.

It is known \cite[Theorem 18.1-1]{Cochran} that (\ref{eq:loveInfty}) has a unique integrable solution $u$ for arbitrary integrable
$g$ if and only if $\Delta_\pm(k)\neq 0$ for $-\infty<k<\infty$.
In our case, we have
\be
  \Kt(0)=\int_{-\infty}^\infty K(x)\,\rd x=1, \label{KtZERO}
\ee
which means $\Delta_-(0)=0$. Hence, we must distinguish
(${\rm I}_g^+$) and (${\rm I}_g^-$). We start with the uniquely-solvable equation,
(${\rm I}_g^+$), in \ref{Igpsec}, and then discuss
(${\rm I}_g^-$) in \ref{Igmsec}.

We note that much of the classical theory makes use of analytic function theory \cite[section~8.5]{MF}.
However, this approach does not seem to be applicable here.
The integral defining $\Kt(k)$ diverges for non-real $k$. Moreover,  $\Kt(k)=\ce^{-\alpha k}$ for $k>0$
continues analytically for all complex $k$ but this continuation does not agree with 
the known formula when $k$ is real and negative, $\Kt(k)=\ce^{\alpha k}$.

%%%%%%%%%%%%%%%%%
\subsection{Equation (${\rm I}_g^+$)}\label{Igpsec}

Start with the formula (\ref{KtZERO}). It shows that
\be
\mbox{$u(x)=\frac{1}{2}$ solves (${\rm I}_g^+$) when $g(x)=1$.}\label{gONE}
\ee
 This $g$ and the solution $u$ do not have Fourier transforms.

\subsubsection{Three examples}

We give three examples, with simple Fourier-trans\-form\-able~$g$.

Motivated by (\ref{gONE}), suppose
$g$ is a piecewise-constant even function defined by
\be
   g(x)=\left\{\begin{array}{ll}
   1,& |x|<L,\\
   0, & |x|>L,
   \end{array}\right. \label{tophatg}
\ee
for some $L>0$. Then    $\gt(k)=(2/k) \sin(kL)$
and, inverting (\ref{utgt}),
\begin{align*}
  u(x)&=\frac{1}{2\pi}\int_{-\infty}^\infty  \frac{2\sin(kL)}{k(1+\ce^{-\alpha|k|})}\,
  \ce^{-\ci kx}\,\rd k\\
  &
  =\frac{2}{\pi}\int_0^\infty \frac{\sin(kL)\cos(kx)}{k(1+\ce^{-\alpha k})}\,\rd k
  =S(L+x)+S(L-x),
\end{align*}
say, where
\[
S(X)=  
  \frac{1}{\pi}\int_0^\infty \frac{\sin(kX)}{k(1+\ce^{-\alpha k})}\,\rd k.
\]
Although we have been unable to evaluate $S(X)$ explicitly, let us
find an asymptotic approximation as $X\to\infty$.
As $1+\ce^{-\alpha k}\sim 2$ as $k\to 0$, write
\begin{align*}
  S(X)&=  
  \frac{1}{\pi}\int_0^\infty \frac{\sin(kX)}{2k}\,\rd k
+\frac{1}{\pi}\int_0^\infty \frac{(1-\ce^{-\alpha k})\sin(kX)}{ 2k(1+\ce^{-\alpha k})}\,\rd k\\
&=\frac{1}{4} +\frac{1}{2\pi}\,\Imag\int_0^\infty \Phi(k)\,\ce^{\ci kX}\,\rd k,
\end{align*}
with $\Phi(k)=k^{-1}\tanh{(\alpha k/2)}$. We have $\Phi(k)\to\alpha/2$ as $k\to 0$
and $\Phi(k)\to 0$ as $k\to\infty$. Then a standard argument for estimating
Fourier integrals \cite[section~10]{Copson} (essentially using integration by parts)
gives
\[
  S(X)\sim \frac{1}{4} +\frac{\alpha}{4\pi X} \quad\mbox{as $X\to\infty$}.
\]
Hence
\[
  u(x)\sim\frac{1}{2}+\frac{\alpha}{2\pi L}\quad
  \mbox{as $L\to\infty$, for fixed~$x$.}
\]
Thus, we recover the known solution for $g(x)=1$, (\ref{gONE}),  as $L\to\infty$;
see~(\ref{tophatg}).

Suppose $g$ is a smooth odd function defined by
\be
  g(x)=\frac{x}{x^2+\kappa^2} \quad\mbox{with}\quad
  \gt(k)=\pi\ci \,\ce^{-\kappa |k|}\,{\rm sgn}\,(k), \label{ETodd}
\ee
where $\kappa>0$.
Inverting (\ref{utgt}),
\begin{align}
  u(x)&=\frac{1}{2\pi}\int_{-\infty}^\infty \frac{\pi\ci \,\ce^{-\kappa|k|}\,{\rm sgn}\,(k)}{1+\ce^{-\alpha|k|}}\,\ce^{-\ci kx}\,\rd k
  =\int_0^\infty \frac{\ce^{-\kappa k}\sin(kx)}{1+\ce^{-\alpha k}}\,\rd k\label{uEx2INT}\\
  &=\frac{1}{\alpha} \,\Imag\int_0^\infty \frac{\ce^{-Z y}\,\rd y}{1+\ce^{-y}}
  =\frac{1}{\alpha} \,\Imag\left\{\beta(Z)\right\},
\end{align}
using \cite[8.371.2]{GR}, where $Z=(\kappa-\ci x)/\alpha$,
\be
  \beta(z)=\frac{1}{2}\left\{ \psi\left(\frac{z+1}{2}\right) -\psi\left(\frac{z}{2}\right)\right\},
  \quad
  \psi(z)=\frac{\Gamma'(z)}{\Gamma(z)} \label{betaDEFN}
\ee
and $\Gamma$ is the gamma function.

When $\alpha=\kappa$, 
the integral (\ref{uEx2INT}) can be evaluated explicitly~\cite[3.911.1]{GR}:
\[
  u(x)=\frac{1}{2x}-\frac{\pi}{2\alpha\sinh{(\pi x/\alpha)}},\quad x>0,
\]
with $u(x)=-u(-x)$ for $x<0$.
Note that, although this solution for $u$ is Fourier-trans\-form\-able, it is not absolutely integrable.
(There is a  similar example for (${\rm I}_g^-$) in \cite{ECT}; see (\ref{ECTex}) below.)

Consider an even version of (\ref{ETodd}),
\be
  g(x)=\frac{\kappa}{x^2+\kappa^2} \quad\mbox{with}\quad
  \gt(k)=\pi \,\ce^{-\kappa |k|}, \label{ETeven}
\ee
where $\kappa>0$. Hence, proceeding as with (\ref{ETodd}),
\be
  u(x)=  \int_0^\infty \frac{\ce^{-\kappa k}\cos(kx)}{1+\ce^{-\alpha k}}\,\rd k
   =\frac{1}{\alpha} \,\Real\left\{\beta(Z)\right\}. \label{uEx3INT}
\ee

When $\alpha=2\kappa$, the integral (\ref{uEx3INT}) can be evaluated explicitly
\cite[3.981.3]{GR}:
\[
  u(x)=\frac{\pi}{2\alpha}\,{\rm sech}{\left(\frac{\pi x}{\alpha}\right)}.
\]
In particular, when $\alpha=2$ ($\kappa=1$), we recover a solution of 
 (${\rm I}_g^+$) found by Hulth\'en \cite[\eqn~(III,\,56)]{Hulthen38};
see also \cite[\eqn~(19)]{Griffiths1964}.

\subsubsection{The resolvent kernel $M_+$}

Next, let us return to (\ref{ugMCT}) and  evaluate the resolvent kernel $M_+(x)$. We have
\begin{align*}
  M_+(x)&=\frac{1}{2\pi}\int_{-\infty}^\infty \frac{\ce^{-\alpha|k|}}{1+\ce^{-\alpha|k|}}\,\ce^{-\ci kx}\,\rd k
  =\frac{1}{\pi}\int_0^\infty \frac{\ce^{-\alpha k}\cos(kx)}{1+\ce^{-\alpha k}}\,\rd k\\
  &=\frac{1}{\pi\alpha}\,\Real\int_0^\infty \frac{\ce^{-\mu y}\,\rd y}{1+\ce^{-y}}
    =\frac{1}{\pi\alpha} \,\Real\left\{\beta(\mu)\right\}
    \quad\mbox{with}\quad \mu=1+\frac{\ci x}{\alpha},
\end{align*}
where $\beta(z)$ is defined by~(\ref{betaDEFN}).

%%%%%%%%%%%%%%%%%
\subsection{Equation (${\rm I}_g^-$)}\label{Igmsec}

The formula (\ref{KtZERO}) implies that the homogeneous form of
 (${\rm I}_g^-$) is satisfied by $u(x)=1$, so that we do not have uniqueness.
We could restore uniqueness by insisting that $u$ be integrable.
Alternatively,  when $g$ is odd we could insist that the solution $u$ be odd.

As an example with an odd $g$, take (\ref{ETodd}). Then (\ref{utgt}) gives
\[
 u(x)  =\int_0^\infty \frac{\ce^{-\kappa k}\sin(kx)}{1-\ce^{-\alpha k}}\,\rd k.
\]
We see that both numerator and denominator are zero at $k=0$ with a finite ratio, and so the integral is well defined.
Indeed, from \cite[3.911.6]{GR}, we have
\[
  u(x)
  =-\frac{1}{\alpha}\,\Imag\left\{\psi(Z)\right\},
\]
where $Z=(\kappa-\ci x)/\alpha$ (as before).
In the special case $\alpha=\kappa$, we have \cite[3.911.2]{GR} 
\be
  u(x)=\frac{\pi}{2\alpha}\coth\left(\frac{\pi x}{\alpha}\right)-\frac{1}{2x}, \label{ECTex}
\ee
in agreement with an example in Titchmarsh's book \cite[p.~309]{ECT}.

For an even example, take (\ref{ETeven}).
Inverting (\ref{utgt}) gives
\[
   u(x)=\frac{1}{2}\int_{-\infty}^\infty \frac{\ce^{-\kappa |k|}\,\ce^{-\ci kx}}{1-\ce^{-\alpha |k|}}\,\rd k
   =\int_0^\infty  \frac{\ce^{-\kappa k}\cos(kx) }{1-\ce^{-\alpha k}}\,\rd k.
\]
The integrand has a non-integrable singularity at $k=0$.
We take the finite part, and define
\[
  \newint_0^\infty  \frac{\ce^{-\kappa k}\cos(kx) }{1-\ce^{-\alpha k}}\,\rd k
   =\lim_{\ve\to 0}\left\{\int_\ve^\infty  \frac{\ce^{-\kappa k}\cos(kx) }{1-\ce^{-\alpha k}}\,\rd k
   +\frac{1}{\alpha}\log(\ve)\right\}.
\]
More generally, define
\[  
  \newint_0^A  \frac{G(k)\cos(kx) }{1-\ce^{-\alpha k}}\,\rd k
   =\lim_{\ve\to 0}\left\{\int_\ve^A  \frac{G(k)\cos(kx) }{1-\ce^{-\alpha k}}\,\rd k
   +\frac{G(0)}{\alpha}\log(\ve)\right\}.
\]
Notice that the second term on the right-hand side of this formula does not depend on~$x$.
However, we are not interested in additive constants because we already know that $u=1$ 
solves the homogeneous version of~(${\rm I}_g^-$).

Let us write
\[
  \gt(k)=\gte(k)+\gto(k)
\]
where $\gte(-k)=\gte(k)$ and $\gto(-k)=-\gto(k)$. Then
\[
  u(x)=\frac{1}{\pi}\newint_0^\infty \frac{\gte(k)\cos(kx)}{1-\ce^{-\alpha k}}\,\rd k
  -\frac{\ci}{\pi}\int_0^\infty \frac{\gto(k)\sin(kx)}{1-\ce^{-\alpha k}}\,\rd k
\]
is a particular solution of~(${\rm I}_g^-$).

%%%%%%%%%%%%%%%%%%%%%%%%%%%%%%%%%%%%%%%%%%%%%%%%%%%%%%%%%%%%%%%%%%%%%%%%%%%%%%%%%%%%%%%

\end{document}